\documentclass[a4paper,12pt]{article}

\usepackage[usenames]{color}
\usepackage{graphicx}
\usepackage{amsmath}
\usepackage{amssymb}

\makeatletter
\@addtoreset{equation}{section}
\makeatother

\newcommand{\be}{\begin{equation}}
\newcommand{\ee}{\end{equation}}
\newcommand{\bea}{\begin{eqnarray}}
\newcommand{\eea}{\end{eqnarray}}

\def\double #1{#1{\hbox{\kern-2pt $#1$}}}

\begin{document}
\begin{titlepage}
\null
\begin{flushright}
December, 2012
\end{flushright}

\vskip 1.8cm
\begin{center}
 
  {\Large \bf Non-Abelian Gauge Groups and Hypermultiplets \\
 \vskip 0.2cm 
in Projective Superspaces
}

\vskip 1.8cm
\normalsize

  {\bf Masato Arai$^{\dagger}$\footnote{masato.arai(at)fukushima-nct.ac.jp} 
and Shin Sasaki$^\sharp$\footnote{shin-s(at)kitasato-u.ac.jp}}

\vskip 0.5cm

  { \it 
  $^\dagger$Fukushima National College of Technology \\
   Iwaki, Fukushima 970-8034, Japan \\
  \vskip 0.5cm
  $^\dagger$Institute of Experimental and Applied Physics \\
Czech Technical University in Prague \\
Horsk\' a 3a/22, 128 00 Prague 2, Czech Republic \\
    \vskip 0.5cm
  $^\sharp$Department of Physics \\
  Kitasato University \\
  Sagamihara, 252-0373, Japan
  }

\vskip 2cm

\begin{abstract}
We construct off-shell superconformal actions of 
hypermultiplets coupled with non-Abelian gauge multiplets in 
 three-dimensional $\mathcal{N} = 3$ and $\mathcal{N} = 4$ 
 projective superspaces. 
We establish the explicit embeddings of 
the $\mathcal{N} = 2$ vector and adjoint chiral superfields,
that constitute the $\mathcal{N} = 4$ gauge multiplets, into the 
tropical multiplets.
We also construct the action in the four-dimensional $\mathcal{N} = 2$
 projective superspace.
\end{abstract}

\end{center}
\end{titlepage}

\newpage

\section{Introduction}
Gauge theories with extended supersymmetries have been studied in
various contexts. 
Among them, superspace formalism of gauge theories is quite useful 
since supersymmetries of an action is manifestly guaranteed.
However the standard superfield formalism is not suitable for theories with extended
supersymmetries. 
The projective superspace formalism is one of the 
useful frameworks to treat the extended supersymmetries.
The four-dimensional $\mathcal{N} = 2$ projective superspace was
established in \cite{KaLiRo, GaHuRo, LiRo1}. 
The projective superspace formalisms were studied 
in two \cite{2d}, three \cite{KuPaTaUn}, five and six dimensions \cite{KuLi, 6d}.

In four-dimensional $\mathcal{N} = 2$ projective superspace, 
a non-Abelian gauge multiplet is introduced as a tropical multiplet.
The kinetic term of the gauge multiplet is defined through the
anti-commutation relations of the gauge covariantized supercovariant
derivatives \cite{LiRo2}.
The formal treatments of interactions among
 hypermultiplets and non-Abelian gauge fields are analyzed in \cite{Go}.
On the other hand, an $\mathcal{N} = 2$ gauge multiplet consists of
an $\mathcal{N} = 1$ vector and an adjoint chiral superfields.
Consequently, the component superfields in the tropical multiplet should be
represented by the $\mathcal{N} = 1$ vector and the adjoint chiral superfields.
However, the explicit embeddings of the tropical multiplet into 
the gauge multiplet remain less understood. 
In the previous paper, 
we have constructed superconformal 
Abelian Chern-Simons matter action in the three-dimensional 
$\mathcal{N} = 3$ and $\mathcal{N} = 4$ projective
superspaces \cite{ArSa}. 
We have shown the explicit relations among
 components in the tropical multiplet and the superfields in the Abelian
 gauge multiplet.
For non-Abelian gauge groups, the relation between the tropical
and gauge multiplets becomes highly non-linear and non-trivial.

The purpose of this paper is to find explicit relations among component
superfields in the tropical multiplet and superfields in the 
non-Abelian gauge multiplet in three and four dimensions.
We will write down the gauge connections in terms of the tropical
multiplet. The vector and chiral superfields in the non-Abelian gauge multiplet are
explicitly expressed by the components of the tropical multiplet. 
We will then show that the adjoint chiral superfields obtained from the
tropical multiplet have correct chiral properties.
We also write down actions of 
hypermultiplets in three-dimensional $\mathcal{N} = 3$, $\mathcal{N} =
4$ and four-dimensional $\mathcal{N} = 2$ superspaces 
in terms of the component superfields in the gauge multiplet. 
In three-dimensions, the action have superconformal invariance. 

The organization of this paper is as follows. 
In the next section, we give a brief review of the three-dimensional
$\mathcal{N} = 3$ and $\mathcal{N} = 4$ 
projective superspace formulations of superconformal theories.
We then introduce the four-dimensional $\mathcal{N} = 2$ formalism.
In Section 3, we construct the action of hypermultiplets coupled with 
 non-Abelian gauge multiplets.
We explicitly write down the gauge connection in terms of the tropical
multiplet. We find perturbative expressions of the adjoint scalar fields
in the vector multiplets and show its chiralities.
We then show that the actions constructed in the projective superspaces 
precisely reproduce those in the $\mathcal{N} = 2$ subsuperspace in three
dimensions. 
We also construct actions in four dimensions.
Section 4 is conclusion and discussions.
Notations and conventions of three-dimensional $\mathcal{N} = 2,3,4$ 
and four-dimensional $\mathcal{N} = 1,2$ superspaces are given in
Appendix A. 
Explicit expressions for the decomposition of the tropical multiplet are 
found in Appendix B.
The anti-commutation relations among the 
gauge covariantized supercovariant derivatives are found in Appendix C.

\section{Projective superspaces}
In this section, we give a brief overview of the projective superspace
formalism.
We first explain the basic facts about 
the three-dimensional $\mathcal{N} = 3$ projective superspace formalism.
We introduce the general formula of actions for $\mathcal{N}=3$ superconformal theories.
Then we generalize the formalism to $\mathcal{N} = 4$ theories.
We also introduce the four-dimensional $\mathcal{N} = 2$ projective
superspace formalism.

\subsection{$\mathcal{N} = 3$ superconformal theories in three-dimensional projective superspace}
The $\mathcal{N} = 3$ projective superspace are defined by the 
standard $\mathcal{N} = 3$ superspace $\mathbb{M}^{3|6}$ 
supplemented with the internal space $\mathbb{C}P^1$.
The coordinates in each space are $z^M = (x^{m},\theta^{\alpha}_I)$ and 
$SU(2)_R$ complex isospinors $v^i, u^i$.
Here $\alpha=1,2$ is the 
$SO(1,2) \sim SL(2,\mathbb{R})$ Lorentz spinor and $I = 1,2, 3$ is 
the $SO(3)_R \sim SU(2)_R$ R-symmetry vector 
index respectively. 
The two complex isospinors satisfy the following
completeness relation, 
\begin{eqnarray}
\delta^i {}_j = \frac{1}{(v,u)} (v^i u_j - v_j u^i), \quad
(v,u) \equiv v^i u_i,
\label{completeness}
\end{eqnarray}
where $u_i$ should obey the condition $(v,u) \not= 0$.
In the following, we use the $SU(2)_R$ spinor indices $i,j = 1,2$ rather than the $SO(3)_R$
vector indices. These are intertwined by the relation
$\theta^{\alpha}_{ij} = (\tau_I)_{ij} \theta^{\alpha}_I$ where
$(\tau_I)^i {}_j$ are the Pauli matrices.
The $SU(2)_R$ indices are raised and
lowered by the anti-symmetric symbols $\varepsilon^{ij}, \varepsilon_{ij}$
such as $\theta^i = \varepsilon^{ij} \theta_j$. 

Using the isospinors $v^i, u^i$, 
we define the following supercovariant derivatives in the projective
superspace, 
\begin{eqnarray}
D^{(2)}_{\alpha} = v_i v_j D^{ij}_{\alpha}, \quad
D^{(0)}_{\alpha} = \frac{1}{(v,u)} v_i u_j D^{ij}_{\alpha}, \quad 
D^{(-2)}_{\alpha} = \frac{1}{(v,u)^2} u_i u_j D^{ij}_{\alpha}.
\end{eqnarray}
The covariant derivative $D^{(2)}_\alpha$ is used to 
define a superconformal projective multiplet.
We define a projective superfield $Q^{(n)}$ by the following constraint:
\begin{eqnarray}
D^{(2)}_{\alpha} Q^{(n)} = 0.
\label{Sconstraint} 
\end{eqnarray}
The superfield $Q^{(n)}$ should be holomorphic with respect to 
$v^i$ and homogeneous function of degree $n$,
\begin{eqnarray}
Q^{(n)} (z, cv) = c^n Q^{(n)} (z, v), \quad c \in \mathbb{C}^{*}.
\label{homogeneous}
\end{eqnarray}
We call $Q^{(n)}$ superconformal projective multiplet with weight 
$n \in \mathbb{Z}$.
The $\mathcal{N} = 3$ superconformal transformation of the superfield
$Q^{(n)}$ is given by \cite{KuPaTaUn},
\begin{eqnarray}
 \delta Q^{(n)}=-(\xi-\Lambda^{(2)}
\boldsymbol{\partial}^{(-2)})Q^{(n)}-n\Sigma Q^{(n)},
\end{eqnarray}
where $\xi = \xi^m \partial_m + \xi^{\alpha}_I D^I_{\alpha}$ with 
$D_\alpha^{ij}=(\tau_I)^{ij}D_\alpha^I$ 
is the superconformal killing vector field.
The symbol $\boldsymbol{\partial}^{(-2)}={1 \over (v,u)}u^i{\partial \over \partial
v^i}$ stands for the differentiation with respect to the isospinor $v^i$.
The parameters $\Lambda^{(2)}=v_iv_j\Lambda^{ij}$ 
and $\Sigma$ correspond to the $SO(3)_R$ and the scale transformations 
respectively. 

The ``smile conjugation'' which is consistent with the constraint
\eqref{Sconstraint} is defined by 
\begin{eqnarray}
\bar{Q}^{(n)} (v) \equiv \left. \left({Q^{(n)} (v)}\right)^*  \right|_{(v^i)^* \to -v_i},
\end{eqnarray}
where the symbol ``$*$'' stands for the ordinary complex conjugation.
The replacement of $v^i$ is explicitly given by 
$(v^1)^* \rightarrow -v_1 = v^2$ and $(v^2)^* \rightarrow
-v_2 = -v^1$.
When the weight $n$ is even number, one can define a real projective
multiplet by using the smile conjugation.

The $\mathcal{N} = 3$ superconformal invariant action is given by \cite{KuPaTaUn}
\begin{eqnarray}
S = 
 \frac{1}{8\pi} \oint_{\gamma} \!
(v, dv) 
 \int \! d^3 x 
\ (D^{(-2)})^2 (D^{(0)})^2 \left. \mathcal{L}^{(2)} (z,v)
 \right|_{\theta = 0},
\label{action}
\end{eqnarray}
where $\mathcal{L}^{(2)}$ is an weight-2 real superconformal projective multiplet.
For later convenience we call $\mathcal{L}^{(2)}$ Lagrangian.
The line integral is evaluated over a closed contour $\gamma$ in $\mathbb{C}P^1$. 
The action \eqref{action} is independent of $u$ and 
we can choose a frame where $u_i = (1,0)$.

The action \eqref{action} is rewritten in terms of 
$\mathcal{N}=2$ superspace and superfields as follows.
We first take the contour $\gamma$ in \eqref{action} such that 
it does not pass through the north pole $v^i=(0,1)$ 
in $\mathbb{C}P^1$.
It is then useful to introduce a complex inhomogeneous coordinate 
$\zeta \in \mathbb{C}$ in the upper hemisphere of $\mathbb{C}P^1$,
\begin{eqnarray}
v^i = v^1 (1, \zeta), \quad \zeta \equiv \frac{v^2}{v^1}, \quad i=1,2.
\end{eqnarray}
Then the supercovariant derivative $D^{(2)}_{\alpha}$ turns into the
following form
\begin{eqnarray}
 D^{(2)}_{\alpha} = (v^1)^2 D^{[2]}_{\alpha},  \qquad 
  D^{[2]}_{\alpha} (\zeta) \equiv - \bar{\mathbb{D}}_{\alpha} - 2 \zeta
 D^{12}_{\alpha} + \zeta^2 \mathbb{D}_{\alpha}, 
\label{cov2}
\end{eqnarray}
where we have introduced the $\mathcal{N} = 2$ supercovariant derivatives
$\mathbb{D}_{\alpha}$ and $\bar{\mathbb{D}}_{\alpha}$ (see Appendix A).
By factoring out the $v^1$ dependence in $Q^{(n)} (z,v)$, 
a new superfield $Q^{[n]} (z,v)\propto Q^{(n)} (z,v)$ 
which satisfies the following constraint is defined:
\begin{eqnarray}
D^{[2]}_{\alpha} (\zeta) Q^{[n]} (z, \zeta) = 0. 
\label{Pconstraint}
\end{eqnarray}
In general, $Q^{[n]}$ is expanded by power series 
in $\zeta$,
\begin{eqnarray}
Q^{[n]} (z, \zeta) = \sum_k \zeta^k Q_k (z),
\end{eqnarray}
where $Q_k (z)$ are standard $\mathcal{N} = 3$ superfields subject
to the constraints \eqref{Pconstraint}.
By the factorization of $v^1$, the Lagrangian $\mathcal{L}^{(2)}$ is rewritten as
\begin{eqnarray}
\mathcal{L}^{(2)} (z, v) = (v^1)^2 (i\zeta) \mathcal{L}^{[2]} (z, \zeta).
\end{eqnarray}
Then the action \eqref{action} reduces to the following form,
\begin{eqnarray}
S = \frac{1}{2\pi i} \oint_{\gamma} \! \frac{d\zeta}{\zeta} 
\int d^3 x d^4 \theta \left. \mathcal{L}^{[2]} (z, \zeta)
\right|_{\theta_{12} = 0},
\label{N2action}
\end{eqnarray}
where we have used \eqref{cov2} and the constraint \eqref{Pconstraint}.

We now show a few examples of the projective superfields $Q^{(n)}$. 
\\
\\
\underline{\textbullet \ $\mathcal{O}(k)$ and (ant)arctic multiplets}
\\
\\
The weight-$n$ complex $\mathcal{O}(k)$ multiplet is defined to be holomorphic in the
upper hemisphere in $\mathbb{C}P^1$,
\begin{eqnarray}
\Upsilon^{(n)} (z,v) = (v^1)^n \Upsilon^{[n]} (z, \zeta), \quad 
\Upsilon^{[n]} = \sum^{k}_{l=0} \zeta^l \Upsilon_l (z).
\label{Ok_expansion}
\end{eqnarray}
The constraints \eqref{Pconstraint} on the $\mathcal{N} = 3$ component
superfields are given by 
\begin{eqnarray}
\begin{aligned}
 & \bar{\mathbb{D}}_{\alpha} \Upsilon_0 = 0, \\
 & \bar{\mathbb{D}}_{\alpha} \Upsilon_1 + 2 D^{12}_{\alpha} \Upsilon_0
 = 0, \\
 & \bar{\mathbb{D}}_{\alpha} \Upsilon_l + 2 D_{\alpha}^{12}
 \Upsilon_{l-1} - \mathbb{D}_{\alpha} \Upsilon_{l-2} = 0, \quad 
(2 \le l \le k), \\
 &  2 D^{12}_{\alpha} \Upsilon_k - \mathbb{D}_{\alpha} \Upsilon_{k-1} =
 0, \\
 & \mathbb{D}_{\alpha} \Upsilon_k = 0.
\end{aligned}
\label{Ok_constraint}
\end{eqnarray}
The arctic multiplet is defined as the limit $k \to \infty$ of the 
complex $\mathcal{O}(k)$ multiplet.
The antarctic multiplet is defined by the smile conjugate of the arctic multiplet.
\\
\\
\underline{\textbullet \ $\mathcal{O}(-k,k)$ and tropical multiplets}
\\
\\
The real $\mathcal{O}(-k,k)$ multiplet with weight $2n$ is defined by 
\begin{eqnarray}
\begin{aligned}
&  \mathcal{V}^{(2n)} (z,v) = (i v^1 v^2)^n \mathcal{V}^{[2n]} (z, \zeta) = (v^1)^{2n} (i
 \zeta)^n \mathcal{V}^{[2n]} (z, \zeta),  \\
&  \mathcal{V}^{[2n]} (z, \zeta) = \sum^{k}_{l=-k} \zeta^l V_l (z),
 \quad 
\bar{V}_l = (-1)^l V_{-l},
\label{Okk_expansion}
\end{aligned}
\end{eqnarray}
where the bar in the component superfields represents the ordinary complex conjugate.
The constraints \eqref{Pconstraint} on the $\mathcal{N} = 3$ component
superfields are given by 
\begin{eqnarray}
\begin{aligned}
 & \bar{\mathbb{D}}_{\alpha} V_{-k} = 0, \\
 & \bar{\mathbb{D}}_{\alpha} V_{- k + 1} + 2 D_{\alpha}^{12} V_{-k} = 0, \\
 & \bar{\mathbb{D}}_{\alpha} V_l + 2 D^{12}_{\alpha} V_{l-1} -
 \mathbb{D}_{\alpha} V_{l-2} = 0, \quad 
 (-k+2 \le l \le k), \\
 & 2 D^{12}_{\alpha} V_k - \mathbb{D}_{\alpha} V_{k-1} = 0, \\
 & \mathbb{D}_{\alpha} V_k = 0.
\end{aligned}
\label{Okk_constraint}
\end{eqnarray}
The tropical multiplet is defined as the limit $k \to \infty$ of the 
$\mathcal{O}(-k,k)$ multiplet.

\subsection{$\mathcal{N} = 4$ superconformal theories in
  three-dimensional projective superspace}
The $\mathcal{N} = 4$ superspace $\mathbb{M}^{3|8}$ is parametrized by
the super-coordinates
$
z^M = (x^m, \theta^{\alpha}_{i \bar{j}}),
$
where $i = 1,2$ and $\bar{j} = 1,2$ are indices for the 
$SU(2)_L \times SU(2)_R$ subgroup of $SO(4)_R$ R-symmetry.
These indices are raised and lowered by the antisymmetric matrices
$\varepsilon^{ij}, \varepsilon^{\bar{i} \bar{j}}$ and so on. 
The $\mathcal{N} = 4$ projective superspace is given by 
$\mathbb{M}^{3|8} \times \mathbb{C}P^1_L \times \mathbb{C}P^1_R$. 
We introduce a pair of $\mathbb{C}P^1$ in order to take into 
account the two $SU(2)$ symmetries \cite{KuPaTaUn}.
The complex projective spaces $\mathbb{C}P^1_L \times \mathbb{C}P^1_R$ 
are parametrized by the homogeneous complex coordinates $v_L = (v^i), v_R =
(v^{\bar{k}})$ and $u_L = (u_i), u_R = (u_{\bar{k}})$. 
They satisfy the completeness relation \eqref{completeness} independently.

As in the $\mathcal{N} = 3$ case, we define the following
covariant derivatives:
\begin{eqnarray}
\begin{aligned}
 & D^{(1) \bar{k}}_{\alpha} = v_i D^{i \bar{k}}_{\alpha}, \quad
D^{(-1) \bar{k}}_{\alpha} = \frac{1}{(v_L, u_L)} u_i
D^{i\bar{k}}_{\alpha}, \\
 & D^{(1)i}_{\alpha} = v_{\bar{k}} D^{i \bar{k}}_{\alpha}, \quad 
D^{(-1) i}_{\alpha} = \frac{1}{(v_R, u_R)} u_{\bar{k}} 
D^{i \bar{k}}_{\alpha}.
\end{aligned}
\end{eqnarray}
The supercovariant derivatives $D^{(1){\bar{k}}}_{\alpha},
D^{(-1)\bar{k}}_{\alpha}$ satisfy the following algebras,
\begin{eqnarray}
\begin{aligned}
 & \{ D^{(1) \bar{k}}_{\alpha}, D^{(1)\bar{l}}_{\beta} \} = 
\{ D^{(-1)\bar{k}}_{\alpha}, D^{(-1)\bar{l}}_{\beta} \} = 0, \\
 & \{D^{(1) \bar{k}}_{\alpha}, D^{(-1) \bar{l}}_{\beta}\} = - 2 i
 \varepsilon^{\bar{k} \bar{l}} \partial_{\alpha \beta}. 
\end{aligned}
\end{eqnarray}
The other supercovariant derivatives $D^{(1)i}_{\alpha}, D^{(-1)i}_{\alpha}$ 
satisfy the similar algebras. 
In the $\mathcal{N} = 4$ case, 
associated with two complex projective spaces $\mathbb{C}P^1$, 
one can introduce the left and right weight-$n$ projective
multiplets independently.
They are defined by the following constraints,
\begin{eqnarray}
\begin{aligned}
 & D^{(1) \bar{k}}_{\alpha} Q^{(n)}_L (v_L) = 0, \\
 & D^{(1) i}_{\alpha} Q^{(n)}_R (v_R) = 0. \label{constN4}
\end{aligned}
\end{eqnarray}
Each projective multiplet $Q^{(n)}_L$, $Q^{(n)}_R$ has  
the property \eqref{homogeneous}.
The $\mathcal{N} = 4$ superconformal transformations of the left and
right projective superfields are given by 
\begin{eqnarray}
\begin{aligned}
\delta Q^{(n)}_L =& - \left(
\xi - \Lambda^{(2)}_L 
\boldsymbol{\partial}_L^{(-2)} 
\right) Q^{(n)}_L - n \Sigma_L Q^{(n)}_L, 
\\
\delta Q^{(n)}_R =&  - \left( 
\xi - \Lambda^{(2)}_R 
\boldsymbol{\partial}_R^{(-2)} 
\right) Q^{(n)}_R - n \Sigma_R Q^{(n)}_R,
\end{aligned}
\end{eqnarray}
where $\xi$ is the superconformal Killing vector field, $\Lambda_{L,R}$, 
$\Sigma_{L,R}$ and $\boldsymbol{\partial}^{(-2)}_{L,R}$ 
are defined as in the same way in the $\mathcal{N} = 3$ case \cite{KuPaTaUn}.

In the left part, we introduce the complex inhomogeneous coordinate
$\zeta_L$ by  
\begin{eqnarray}
v^i = v^1 (1, \zeta_L), \quad \zeta_L = \frac{v^2}{v^1}.
\end{eqnarray}
Then the supercovariant derivative in the left part becomes
\begin{eqnarray}
D^{(1)\bar{k}}_{\alpha} = v^1 D^{[1]\bar{k}}_{\alpha}, \qquad 
D^{[1]\bar{k}}_{\alpha} \equiv D_{\alpha}^{2\bar{k}} - \zeta_L D^{1 \bar{k}}_{\alpha}.
\end{eqnarray}
As in the case of the $\mathcal{N} = 3$ formalism, 
the $v^1$ dependencies of the projective superfields 
can be factored out and one can define a new field 
$Q^{[n]}_L \propto Q^{(n)}_L$ which satisfies the following condition,
\begin{equation}
D^{[1]\bar{k}}_{\alpha} (\zeta) Q^{[n]}_L = 0.
\label{N4Lconstraint}
\end{equation}
Therefore the left projective superfield $Q^{[n]}_L$ is expanded as 
\begin{eqnarray}
Q^{[n]}_L (z, \zeta_L) = \sum_k \zeta_L^k Q_k (z),
\end{eqnarray}
where $Q_k(z)$ are the standard $\mathcal{N} = 4$ superfields
subject to the constraint (\ref{constN4}). 
Similar definitions hold in the right part.

The manifestly $\mathcal{N} = 4$ superconformal invariant action 
is given by 
\begin{eqnarray}
S = 
\frac{1}{2\pi} \oint_{\gamma_L} \! 
(v_L, dv_L)
\int \! d^3 x \ D^{(-4)}_L \mathcal{L}_L^{(2)} (z, v_L) |_{\theta = 0} 
+
\frac{1}{2\pi} \oint_{\gamma_R} \! 
(v_R, dv_R)
\int \! d^3 x \ D^{(-4)}_R \mathcal{L}_R^{(2)} (z, v_R) |_{\theta = 0},
\nonumber \\
\label{N4action}
\end{eqnarray}
where $\mathcal{L}^{(2)}_L(\mathcal{L}^{(2)}_R)$ is the weight-2 left (right) projective multiplet and
we have defined the following integration measures,
\begin{eqnarray}
\begin{aligned}
 & D^{(-4)}_L = \frac{1}{48} D^{(-2) \bar{k} \bar{l}} D_{\bar{k}
 \bar{l}}^{(-2)}, \quad D^{(-2)}_{\bar{k} \bar{l}} = D^{(-1)
 \alpha}_{\bar{k}} 
D^{(-1)}_{\alpha \bar{l}}, \\
 & D^{(-4)}_R = \frac{1}{48} D^{(-2)ij} D_{ij}^{(-2)}, \quad
 D^{(-2)}_{ij} = D^{(-1)
 \alpha}_i D^{(-1)}_{\alpha j}.
\end{aligned}
\end{eqnarray}
The contour $\gamma_L$ $(\gamma_R)$ is chosen such that the path goes
the outside of the north pole in $\mathbb{C}P^1_L$ $(\mathbb{C}P^1_R)$.
After fixing $u_i = (1,0)$, $u_{\bar{k}} = (1,0)$ in $\mathbb{C}P^1_L$
and $\mathbb{C}P^1_R$, the action \eqref{N4action} 
is rewritten in the $\mathcal{N} = 2$ superspace:
\begin{eqnarray}
S = 
\frac{1}{2\pi i} \oint_{\gamma_L} \! \frac{d \zeta_L}{\zeta_L} \int
 \! d^3 x d^4 \theta \mathcal{L}^{[2]}_L (z,
 \zeta_L)|_{\theta_{\perp} = 0} 
+ 
\frac{1}{2\pi i} \oint_{\gamma_R} \! \frac{d \zeta_R}{\zeta_R} \int
 \! d^3 x d^4 \theta \mathcal{L}^{[2]}_R (z,
 \zeta_R)|_{\theta_{\perp} = 0},
\end{eqnarray}
where the symbol $|_{\theta_{\perp} = 0}$ means that the 
superfields in the Lagrangian 
are projected on the $\mathcal{N} = 2$ superspace.

Classification of multiplets is similar to the $\mathcal{N} = 3$ case. 
Since the left and right parts have almost the same structure, 
we focus on the left part in the following.
A complex $\mathcal{O}(k)$ multiplet and a real $\mathcal{O}(-k,k)$ multiplet are
defined as (\ref{Ok_expansion}) and (\ref{Okk_expansion}), respectively.
Constraints on the components of a complex
$\mathcal{O}(k)$ multiplet $\Upsilon^{[n]}
= \sum_{l=0}^{k} \Upsilon_l \zeta^l$ 
are given by 
\begin{eqnarray}
\begin{aligned}
 & \bar{\mathbb{D}}_{\alpha} \Upsilon_0 = D_{\alpha}^{2 \bar{1}} \Upsilon_0 = 0, \\
 & \bar{\mathbb{D}}_{\alpha} \Upsilon_l = - D_{\alpha} ^{1 \bar{2}}
 \Upsilon_{l-1}, \quad D_{\alpha}^{2 \bar{1}} \Upsilon_l = \mathbb{D}_{\alpha}
 \Upsilon_{l-1}, \quad (1 \le l \le k), \\
 & \mathbb{D}_{\alpha} \Upsilon_k = D_{\alpha}^{1\bar{2}} \Upsilon_k = 0, 
\end{aligned}
\end{eqnarray}
while those on a real
$\mathcal{O} (-k,k)$ multiplet $\mathcal{V}^{[2n]} = 
\sum_{l=-k}^{k} V_l \zeta^l$ are 
\begin{eqnarray}
\begin{aligned}
 & \bar{\mathbb{D}}_{\alpha} V_{-k} = D_{\alpha}^{2 \bar{1}}
 V_{-k} = 0, \\
 & \bar{\mathbb{D}}_{\alpha} V_l = - D_{\alpha}^{1 \bar{2}} V_{l-1}, \quad
 D_{\alpha}^{2 \bar{1}} V_l = \mathbb{D}_{\alpha}
 V_{l-1},\quad (-k+1 \le l \le k), \\
 & D_{\alpha}^{1 \bar{2}} V_k = \mathbb{D}_{\alpha} V_k = 0.
\end{aligned} \label{G2-constN4}
\end{eqnarray}
The (ant)arctic multiplets and tropical multiplets are defined by 
taking $k\rightarrow \infty$ in the complex $\mathcal{O}(k)$ and the
real $\mathcal{O}(-k,k)$ multiplets, respectively.

\subsection{Four-dimensional $\mathcal{N} = 2$ projective superspace}
The four-dimensional projective superspace is defined by
$\mathbb{M}^{4|8} \times \mathbb{C}P^1$ whose coordinates are 
$(z^M, \zeta) = (x^m, \theta^i_{\alpha}, \bar{\theta}_{i \dot{\alpha}},
\zeta)$ \cite{LiRo1}. 
The $SU(2)_R$ indices $i,j$ run from 1 to 2. $\alpha ,
\dot{\alpha} = 1,2$ are the $SL(2,\mathbb{C})$ Lorentz spinor indices.
A projective multiplet $\Upsilon$ is defined by the following constraints
\begin{eqnarray}
\nabla_{\alpha} \Upsilon = \bar{\nabla}_{\dot{\alpha}} \Upsilon = 0,
\end{eqnarray}
where the operators $\nabla_{\alpha}, \bar{\nabla}_{\dot{\alpha}}$ are
defined by
\begin{equation}
\nabla_{\alpha} = D_{1 \alpha} + \zeta D_{2 \alpha}, \quad 
\bar{\nabla}_{\dot{\alpha}} = \bar{D}^2_{\dot{\alpha}} - \zeta \bar{D}^1_{\dot{\alpha}}.
\end{equation}
The classification of the projective multiplets is the same in the
three-dimensional cases without any weight specified. 
The constraints on the components of the complex $\mathcal{O} (k)$ multiplet
$\Upsilon = \sum_{l=0}^k \Upsilon_l \zeta^l$ are 
\begin{eqnarray}
\begin{array}{l}
  D_{1 \alpha} \Upsilon_0 = 0, \\
  D_{1 \alpha} \Upsilon_l + D_{2 \alpha} \Upsilon_{l-1} = 0, \quad (1 \le
 l \le k), \\
  D_{2 \alpha} \Upsilon_k = 0,
\end{array}
\quad 
\begin{array}{l}
  \bar{D}^2_{\dot{\alpha}} \Upsilon_0 = 0, \\
  \bar{D}^2_{\dot{\alpha}} \Upsilon_l - \bar{D}^1_{\dot{\alpha}}
 \Upsilon_{l-1} = 0, \quad (1 \le l \le k), \\
  \bar{D}^1_{\dot{\alpha}} \Upsilon_k = 0.
\end{array}
\label{eq:4d_const1}
\end{eqnarray}
The constraints on the components of the $\mathcal{O} (-k,k)$ multiplet
$\mathcal{V} = \sum_{l = - \infty}^{\infty} V_l \zeta^l$ are 
\begin{eqnarray}
\begin{array}{l}
  D_{1 \alpha} V_{-k} = 0, \\
  D_{1 \alpha} V_l + D_{2 \alpha} V_{l-1} = 0, 
\quad (-k+1 \le l \le k), 
\\
  D_{2 \alpha} V_{k} = 0, 
\end{array}
\quad 
\begin{array}{l}
  \bar{D}^2_{\dot{\alpha}} V_{-k} = 0, \\
  \bar{D}^2_{\dot{\alpha}} V_l - \bar{D}^1_{\dot{\alpha}} V_{l-1} = 0, 
\quad (-k + 1 \le l \le k), \\
  \bar{D}^1_{\dot{\alpha}} V_k = 0.
\end{array}
\label{eq:4d_const2}
\end{eqnarray}
We introduce the conjugate of the operators $\nabla_{\alpha},
\bar{\nabla}_{\dot{\alpha}}$, 
\begin{eqnarray}
\Delta_{\alpha} = D_{2 \alpha} - \frac{1}{\zeta} D_{1 \alpha}, \quad 
\bar{\Delta}_{\dot{\alpha}} = \bar{D}^1_{\dot{\alpha}} + \frac{1}{\zeta} \bar{D}^2_{\dot{\alpha}}.
\end{eqnarray}
They satisfy the following algebras,
\begin{eqnarray}
\begin{aligned}
 & 
\{\nabla_{\alpha}, \nabla_{\beta} \} = 
\{\nabla_{\alpha}, \bar{\nabla}_{\dot{\alpha}} \} =
\{\bar{\nabla}_{\dot{\alpha}}, \bar{\nabla}_{\dot{\beta}} \} = 0, \\
 & \{\Delta_{\alpha}, \Delta_{\beta}\} = 
\{ \Delta_{\alpha}, \bar{\Delta}_{\dot{\alpha}} \} =
\{ \bar{\Delta}_{\dot{\alpha}}, \bar{\Delta}_{\dot{\beta}} \} = 0, \\
 & \{\nabla_{\alpha}, \Delta_{\beta} \} =
 \{\bar{\Delta}_{\dot{\alpha}}, \bar{\nabla}_{\dot{\beta}} \} = 0, 
\quad  \{\nabla_{\alpha}, \bar{\Delta}_{\dot{\beta}} \} 
= \{\Delta_{\alpha}, \bar{\nabla}_{\dot{\beta}} \} = 2 i
\partial_{\alpha \dot{\beta}}.
\end{aligned}
\end{eqnarray}
The operators $\Delta_{\alpha}, \bar{\Delta}_{\dot{\alpha}}$ are used
to define the integration measure in the four-dimensional $\mathcal{N} = 2$
projective superspace.
The manifestly $\mathcal{N} = 2$ supersymmetry invariant action is given
by 
\begin{eqnarray}
S = \int \! d^4 x \oint_C \! \frac{\zeta d \zeta}{2 \pi i} 
\Delta^2 \bar{\Delta}^2 K,
\end{eqnarray}
where $C$ is a contour surrounding singularities in the $\zeta$-plane and 
$K$ is a gauge invariant function of projective superfields and $\zeta$.

\section{Non-Abelian gauge multiplet and hypermultiplet in projective  superspaces}
In this section we work out the explicit relations between the
non-Abelian gauge multiplet and the tropical multiplet.
In order to find the precise relation, we consider the action of
hypermultiplets coupled with the non-Abelian gauge multiplets.
In terms of the $\mathcal{N} = 2$ language in three dimensions, 
an $\mathcal{N} = 3$ ($\mathcal{N} = 4$) gauge multiplet consists of 
a vector superfield $V$ and adjoint (anti)chiral superfields $\Phi_0,
\bar{\Phi}_0$. 
A hypermultiplet consists of two chiral superfields $S,T$ and whose 
conjugates $\bar{S}, \bar{T}$.
The action of an $\mathcal{N} = 3$ hypermultiplet\footnote{
Strictly speaking, the action \eqref{eq:N3hyper_N2superspace} has
$\mathcal{N} = 4$ supersymmetry in three dimensions. 
Here we keep only $\mathcal{N} = 3$ supersymmetry manifest.
} in the fundamental representation of the
gauge group in $\mathcal{N} = 2$ superspace is 
\begin{eqnarray}
S = \int \! d^3 x d^4 \theta 
\left(
\bar{S} e^{V} S + T e^{-V} \bar{T}
\right)
+ 
\left[
2 \int \! d^3 x d^2 \theta \ T \Phi_0 S
+ 2 \int \! d^3 x d^2 \bar{\theta} \ \bar{T} \bar{\Phi}_0 \bar{S}
\right].
\label{eq:N3hyper_N2superspace}
\end{eqnarray}
The superfields 
$V$, $\Phi_0, \bar{\Phi}_0$ in the action are non-trivial functions of
the component superfields $V_{-1}, V_0, V_1$ in the tropical multiplet. 
We will determine these functions in the following subsections.
The calculation is performed by the perturbation of $V_{-1}, V_0, V_1$.
We work in the three-dimensional $\mathcal{N} = 3$, $\mathcal{N} = 4$
projective superspaces.
In each case, the hypermultiplets are defined by the (ant)arctic
multiplets. 
In four dimensions, the situation is the same in three dimensions.
In the following, we show the detail calculations in three-dimensional
$\mathcal{N} = 3$, $\mathcal{N} = 4$ and four-dimensional $\mathcal{N} =
2$ cases separately.

\subsection{$\mathcal{N} = 3$ in three dimensions}
The gauge multiplet is defined by the weight-0 real tropical multiplet $\mathcal{V}^{[0]}$.
This is adjoint representation of the non-Abelian gauge group $G$.
The gauge transformation of the tropical multiplet is given by 
\begin{eqnarray}
e^{\mathcal{V}^{[0]}} \to e^{- i \bar{\Lambda}^{[0]}}
 e^{\mathcal{V}^{[0]}} e^{i \Lambda^{[0]}},
\label{eq:gauge_transf_tropical}
\end{eqnarray}
where $\Lambda^{[0]}, \bar{\Lambda}^{[0]}$ are weight-0 (ant)arctic
multiplets.
In the Lindstr\"om-Ro\v{c}ek gauge, the tropical multiplet is expanded
as follows \cite{LiRo2},
\begin{eqnarray}
\begin{aligned}
 & \mathcal{V}^{[0]} = \zeta^{-1} V_{-1} + V_0 + \zeta V_1, \\
 & \bar{V}_0 = V_0, \quad \bar{V}_1 = - V_{-1}.
\end{aligned}
\end{eqnarray}
Now we write down the $\mathcal{N} = 3$ gauge multiplet 
in terms of the tropical multiplet $V_{-1}, V_{0}, V_{1}$. 
In the Abelian case, the $\mathcal{N} = 2$ vector superfield $V$ and the
adjoint chiral superfields $\Phi_0, \bar{\Phi}_0$ in the gauge multiplet are
identified as \cite{ArSa}
\begin{equation}
V = V_0, \quad \Phi_0 = \frac{i}{8} \bar{\mathbb{D}}^2 V_1, \quad 
\bar{\Phi}_0 = - \frac{i}{8} \mathbb{D}^2 V_{-1}.
\label{eq:abelian_relation}
\end{equation}
However in the non-Abelian case, the relations 
among $V, \Phi_0, \bar{\Phi}_0$ and $V_{-1}, V_0, V_1$ become non-linear
and the identification is not straightforward. 

We first look for the expression of $V$. 
In order to find it, we decompose $\mathcal{V}^{[0]}$ as follows \cite{LiRo2},
\begin{eqnarray}
e^{\mathcal{V}^{[0]}} \equiv e^{\hat{V}_-} e^{\hat{V}_0} e^{\hat{V}_+},
\label{eq:decomposition}
\end{eqnarray}
where $\hat{V}_{\pm}$ contains terms with positive (negative) powers of $\zeta$
while $\hat{V}_0$ does not depend on $\zeta$.
Following \cite{LiRo2}, we define the gauge transformations of each part:
\begin{eqnarray}
 e^{\hat{V}_-} \to e^{- i \bar{\Lambda}^{[0]}} e^{\hat{V}_-} e^{i
 \bar{\lambda}_0}, \qquad 
 e^{\hat{V}_0} \to e^{- i \bar{\lambda}_0} e^{\hat{V}_0} e^{i
 \lambda_0}, \qquad 
 e^{\hat{V}_+} \to e^{ - i \lambda_0} e^{\hat{V}_+} e^{i
 \Lambda^{[0]}},
\end{eqnarray}
where $\lambda_0, \bar{\lambda}_0$ are $\zeta^0$ components in the
 (ant)arctic multiplets  $\Lambda^{[0]}, \bar{\Lambda}^{[0]}$.
The $\zeta$ independent part $\hat{V}_0$ has correct gauge
 transformation property of the $\mathcal{N} = 2$ vector superfield $V$.
Therefore we identify $\hat{V}_0$ with the vector superfield $V$:
\begin{equation}
V = \hat{V}_0.
\end{equation}
The explicit form of $\hat{V}_0, \hat{V}_{\pm}$ can be calculated
 perturbatively in $V_{-1}, V_0, V_{1}$.
We obtain some explicit expressions at low orders in Appendix B.

Next, we look for the explicit forms of the adjoint chiral superfields 
$\Phi_0, \bar{\Phi}_0$ in terms of
the components in the tropical multiplet.
The adjoint chiral superfields are constructed through the examination of
the interaction terms of hypermultiplets and the non-Abelian gauge multiplet.

The weight-2 Lagrangian of the free hypermultiplet is given by 
\begin{equation}
\mathcal{L}^{[2]} = \bar{\Upsilon}^{[1]} \Upsilon^{[1]}.
\label{eq:free_hyper}
\end{equation}
where $\Upsilon^{[1]}, \bar{\Upsilon}^{[1]}$ are weight-1 arctic and 
antarctic multiplets.
Expanding the projective superfields 
in $\zeta$ and performing the integration over $\zeta$, we
find that the action associated with the Lagrangian \eqref{eq:free_hyper} becomes
\begin{eqnarray}
S = \int \! d^3x d^4 \theta 
\left[
\bar{\Upsilon}_0 \Upsilon_0 - \bar{\Upsilon}_1 \Upsilon_1
+ \mathbb{D}^2 \bar{\Upsilon}_1 Y_0 + \bar{Y}_0 \bar{\mathbb{D}}^2 \Upsilon_1
\right],
\end{eqnarray}
where we have integrated out the auxiliary fields $\Upsilon_{l},
\bar{\Upsilon}_l, \ (l \ge 2)$ and introduced the Lagrange multipliers
$Y_0, \bar{Y}_0$ to impose the constraints of the projective multiplet. 
The components 
$\bar{\Upsilon}_0, \Upsilon_0$ are (anti)chiral superfields in
$\mathcal{N} = 2$ superspace. 
We next integrate out $\bar{\Upsilon}_1, \Upsilon_1$ and dualize these fields
to $\bar{T} \equiv \mathbb{D}^2 Y_0$, $T \equiv \bar{\mathbb{D}}^2
\bar{Y}_0$. The new superfields $\bar{T}, T$ satisfy the (anti)chiral
superfield condition and the action becomes
\begin{equation}
S = \int \! d^3 x d^4 \theta \ 
(\bar{\Upsilon}_0 \Upsilon_0 + \bar{T} T).
\end{equation}
This is nothing but the action of the free hypermultiplet in
$\mathcal{N} = 2$ superspace.
Now we couple the hypermultiplet with the non-Abelian gauge multiplet.
The weight-2 gauge invariant Lagrangian is given by 
\begin{eqnarray}
\mathcal{L}^{[2]} = \bar{\Upsilon}^{[1]} e^{\mathcal{V}^{[0]}}
 \Upsilon^{[1]}.
\label{eq:fundamental_hypermultiplet}
\end{eqnarray}
The gauge transformations of the (ant)arctic multiplets are defined as 
\begin{eqnarray}
\Upsilon^{[1]} \to e^{- i \Lambda^{[0]}} \Upsilon^{[1]}, \quad 
\bar{\Upsilon}^{[1]} \to \bar{\Upsilon}^{[1]} e^{ i \bar{\Lambda}^{[0]}}.
\end{eqnarray}
Especially, for the $\zeta^0$ component we have 
\begin{eqnarray}
\Upsilon_0 \to e^{- i \lambda_0} \Upsilon_0, \quad 
\bar{\Upsilon}_0 \to \bar{\Upsilon}_0 e^{ i \bar{\lambda}_0}.
\end{eqnarray}
We note that these fields have correct gauge transformation of the
$\mathcal{N} = 2$ hypermultiplets (chiral) superfields.
However, writing down the Lagrangian \eqref{eq:fundamental_hypermultiplet}
in terms $\mathcal{N} = 2$ superfields is cumbersome since the tropical
multiplet $\mathcal{V}^{[0]}$ appears in the Lagrangian non-linearly.
In order to find the structure of the Lagrangian in $\mathcal{N} = 2$ superspace,
 we define the following new fields:
\begin{eqnarray}
\bar{\tilde{\Upsilon}}^{[1]} \equiv \bar{\Upsilon}^{[1]} e^{\hat{V}_{-}},
 \qquad 
\tilde{\Upsilon}^{[1]} \equiv e^{\hat{V}_0} e^{\hat{V}_{+}}
\Upsilon^{[1]}.
\label{eq:N3_tilde}
\end{eqnarray}
Then the Lagrangian \eqref{eq:fundamental_hypermultiplet} is rewritten
as the form of the free hypermultiplet:
\begin{eqnarray}
\mathcal{L}^{[2]} = \bar{\tilde{\Upsilon}}^{[1]} \tilde{\Upsilon}^{[1]}.
\end{eqnarray}
The gauge transformation of the new fields are 
\begin{eqnarray}
\bar{\tilde{\Upsilon}}^{[1]} \to \bar{\tilde{\Upsilon}}^{[1]} e^{ i
\bar{\lambda}_0}, \qquad
\tilde{\Upsilon}^{[1]} \to e^{- i \bar{\lambda}_0}
\tilde{\Upsilon}^{[1]}.
\label{eq:tilde_gauge}
\end{eqnarray}
We note that the positive and negative powers of $\zeta$ components in 
$\tilde{\Upsilon}^{[1]},\bar{\tilde{\Upsilon}}^{[1]}$ are never mixed under the
gauge transformations \eqref{eq:tilde_gauge}.

We next specify what constraints the new fields
\eqref{eq:tilde_gauge} satisfy.
The supercovariant derivative $D^{[2]}_{\alpha}$ that 
defines the projective multiplet should be gauge covariantized.
In order that the fields \eqref{eq:N3_tilde} become the covariantly projective
multiplets, we define the following gauge (super)connection,
\begin{equation}
\Omega_{\alpha} \equiv 
e^{ \hat{V}_0} e^{ \hat{V}_{+}} \overrightarrow{D}_{\alpha}^{[2]}
(e^{-  \hat{V}_{+}} e^{-  \hat{V}_0}).
\end{equation}
Then, the left gauge covariant derivative acting on fields with charge $q$ is
defined by 
\begin{eqnarray}
\overrightarrow{\mathcal{D}}_{\alpha}^{[2]} * 
\equiv (e^{q \hat{V}_0} e^{q \hat{V}_{+}}
\overrightarrow{D}^{[2]}_{\alpha} e^{- q \hat{V}_{+}} e^{- q \hat{V}_0}) * 
= \overrightarrow{D}^{[2]}_{\alpha} * 
+ e^{q \hat{V}_0} e^{q \hat{V}_{+}}
\overrightarrow{D}^{[2]}_{\alpha} (e^{- q \hat{V}_{+}} 
e^{- q \hat{V}_{0} }) \triangleright *,
\end{eqnarray}
where the symbol $\triangleright$ means that the quantity acts as the 
appropriate representation of the gauge group $G$.
By the same way, we define the right gauge covariant derivative,
\begin{eqnarray}
* \overleftarrow{\mathcal{D}}_{\alpha}^{[2]} 
\equiv * (e^{ - q \hat{V}_{-}} \overleftarrow{D}^{[2]}_{\alpha} e^{ q
\hat{V}_{-}})
= * \overleftarrow{D}_{\alpha} 
+ * \triangleleft (e^{- q \hat{V}_{-}} \overleftarrow{D}^{(2)}_{\alpha})
e^{q \hat{V}_{-}}.
\end{eqnarray}
We note that the ordering of the product is important since all the
fields are matrix valued in the non-Abelian case.
Then, for the fundamental representation $\tilde{\Upsilon}^{[1]}$,
$\bar{\tilde{\Upsilon}}^{[1]}$ we can show that the new fields
\eqref{eq:N3_tilde} satisfy the gauge covariantized projective constraints:
\begin{eqnarray}
\overrightarrow{\mathcal{D}}_{\alpha}^{[2]} \tilde{\Upsilon}^{[1]} = 0,
\quad 
\bar{\tilde{\Upsilon}}^{[1]} 
\overleftarrow{\mathcal{D}}_{\alpha}^{[2]} = 0.
\end{eqnarray}
The gauge transformation of the connection is 
\begin{eqnarray}
\Omega_{\alpha} \to \Omega_{\alpha}' 
= e^{- i q \bar{\lambda}_0} \Omega_{\alpha} e^{i q \bar{\lambda}_0} +
e^{- i q \bar{\lambda}_0} (D_{\alpha}^{[2]} e^{i q \bar{\lambda}_0}).
\end{eqnarray}
This is the typical gauge transformation associated with the 
transformation \eqref{eq:tilde_gauge}.

We next calculate the gauge connection $\Omega_{\alpha}$.
From the constraint condition $D_{\alpha}^{[2]} \mathcal{V}^{[0]} = 0$,
we have an identity, 
\begin{eqnarray}
e^{- \hat{V}_-} (D^{[2]}_{\alpha} e^{\hat{V}_-}) = e^{\hat{V}_0}
 e^{\hat{V}_+} D^{[2]}_{\alpha} (e^{- \hat{V}_+} e^{- \hat{V}_0}).
\label{eq:ident}
\end{eqnarray}
Using the $\zeta$-expansion of the supercovariant derivative 
$D^{[2]}_{\alpha} = -
\bar{\mathbb{D}}_{\alpha} - 2 \zeta D^{12}_{\alpha} + \zeta^2
\mathbb{D}_{\alpha}$, the both sides in \eqref{eq:ident} are expanded as 
\begin{eqnarray}
& & e^{- \hat{V}_{-}} (D^{[2]}_{\alpha} e^{\hat{V}_{-}}) 
\equiv  \zeta^2 \Gamma^{(-)}_{-2 \alpha} + \zeta
 \Gamma^{(-)}_{-1 \alpha} + \zeta^0 \Gamma^{(-)}_{0 \alpha} + \sum_{l=1}^{\infty}
 \zeta^{-l} \Gamma^{(-)}_{l \alpha}, \\
& & e^{\hat{V}_0}
 e^{\hat{V}_+} D^{[2]}_{\alpha} (e^{- \hat{V}_+} e^{- \hat{V}_0}) 
\equiv \sum_{l = 0}^{\infty} \zeta^l
 \Gamma^{(+)}_{l \alpha}.
\end{eqnarray}
Then we have the following relations among the components,
\begin{eqnarray}
\begin{aligned}
 & \Gamma^{(+)}_{2 \alpha} = \Gamma^{(-)}_{-2 \alpha}, \quad 
\Gamma^{(+)}_{1 \alpha} = \Gamma^{(-)}_{-1 \alpha}, \quad 
\Gamma^{(+)}_{0 \alpha} = \Gamma^{(-)}_{0 \alpha}, \\
 & \Gamma^{(+)}_{l \ge 3 \ \alpha} = \Gamma^{(-)}_{l \ge 1 \ \alpha} =
 0.
\end{aligned}
\end{eqnarray}
The gauge connection is therefore expressed as 
\begin{equation}
\Omega_{\alpha} = 
\zeta^{2} \Gamma^{(-)}_{-2 \alpha} + \zeta
 \Gamma^{(-)}_{-1 \alpha} + \zeta^0 \Gamma^{(-)}_{0 \alpha}.
\end{equation}
Consequently the gauge covariantized supercovariant derivative contains $\zeta^0, \zeta, \zeta^2$
components:
\begin{equation}
\mathcal{D}^{[2]}_{\alpha} = D^{[2]}_{\alpha} +
\Omega_{\alpha} \equiv - \bar{\mathcal{D}}_{\alpha} - 2 \zeta
\mathcal{D}^{12}_{\alpha} + \zeta^2 \mathcal{D}_{\alpha},
\end{equation}
where we have defined the gauge covariantized supercovariant derivatives
\begin{eqnarray}
\mathcal{D}_{\alpha} = \mathbb{D}_{\alpha} + \Gamma^{(-)}_{2
 \alpha}, \qquad 
\mathcal{D}^{12}_{\alpha} = D^{12}_{\alpha} - \frac{1}{2}
 \Gamma^{(-)}_{-1 \alpha}, \qquad 
\bar{\mathcal{D}}_{\alpha} = \bar{\mathbb{D}}_{\alpha} -
 \Gamma^{(-)}_{0 \alpha}.
\end{eqnarray}
The anti-commutation relations of the above gauge covariant derivatives are found in 
Appendix C.
From the explicit form of the left hand side in \eqref{eq:ident}
we find that the $\zeta^2$ term in the gauge connection vanishes identically,
\begin{equation}
\Gamma^{(-)}_{-2 \alpha} = 0.
\label{eq:G2_full}
\end{equation}
In general, $\hat{V}_{-}$ in the left hand side of \eqref{eq:decomposition}
is expanded as 
$\hat{V}_{-} = \sum_{l=1}^{\infty} \zeta^{-l} \hat{V}_{-l}$. 
Then we find 
\begin{eqnarray}
e^{- \hat{V}_{-}} D^{[2]}_{\alpha} e^{\hat{V}_{-}} 
= 
- 2 D^{12}_{\alpha} \hat{V}_{-1} + \mathbb{D}_{\alpha} \hat{V}_{-2} 
+ \frac{1}{2} [\mathbb{D}_{\alpha} \hat{V}_{-1}, \hat{V}_{-1}] + \zeta
\mathbb{D}_{\alpha} \hat{V}_{-1}.
\end{eqnarray}
Therefore we obtain the following form of the components in the gauge
connection,
\begin{eqnarray}
\begin{aligned}
\Gamma^{(-)}_{-2 \alpha} &= 0, \\
\Gamma^{(-)}_{-1 \alpha} &= \mathbb{D}_{\alpha} \hat{V}_{-1}
\\
\Gamma^{(-)}_{0 \alpha} &= 
\mathbb{D}_{\alpha} \hat{V}_{-2} - 2 D^{12}_{\alpha} \hat{V}_{-1} 
+ \frac{1}{2} [\mathbb{D}_{\alpha} \hat{V}_{-1}, \hat{V}_{-1}].
\end{aligned}
\end{eqnarray}
Only the $\zeta^{-1}, \zeta^{-2}$ components in $\hat{V}_{-}$ contribute
to the gauge connection.
The explicit representations of $\hat{V}_{-1}, \hat{V}_{-2}$ in the decomposition of 
$e^{\mathcal{V}^{[0]}}$ 
are calculated in Appendix B.

The constraints on the component superfields in the 
(ant)arctic multiplet \eqref{eq:N3_tilde} are 
now gauge covariantized:
\begin{eqnarray}
\begin{aligned}
 & \bar{\mathcal{D}}_{\alpha} \tilde{\Upsilon}_0 = 0, \\
 & \bar{\mathcal{D}}_{\alpha} \tilde{\Upsilon}_1 + 2 \mathcal{D}^{12}_{\alpha} \tilde{\Upsilon}_0
 = 0, \\
 & \bar{\mathcal{D}}_{\alpha} \tilde{\Upsilon}_{l} + 2 \mathcal{D}^{12}_{\alpha}
 \tilde{\Upsilon}_{l-1} - \mathcal{D}_{\alpha} \tilde{\Upsilon}_{l-2} = 0, \quad (l \ge
 2), \\
 & \bar{\tilde{\Upsilon}}_0 \overleftarrow{\mathcal{D}}_{\alpha} = 0, \\
 & \bar{\tilde{\Upsilon}}_1 \overleftarrow{\mathcal{D}}_{\alpha} 
- 2 \bar{\tilde{\Upsilon}}_0 \overleftarrow{\mathcal{D}}^{12}_{\alpha} = 0, \\
 & \bar{\tilde{\Upsilon}}_{l-2} \overleftarrow{\mathcal{D}}_{\alpha}
+ 2 \bar{\tilde{\Upsilon}}_{l-1} \overleftarrow{\mathcal{D}}^{12}_{\alpha}
- \bar{\tilde{\Upsilon}}_{l} \overleftarrow{\mathcal{D}}_{\alpha} = 0,
 \quad (l \ge 2).
\end{aligned}
\end{eqnarray}
From the above constraints 
and the anti-commutation relations of the gauge covariantized
supercovariant derivatives in Appendix C, 
we find the following relations,
\begin{eqnarray}
\begin{aligned}
 & \bar{\mathcal{D}}^2 \tilde{\Upsilon}_1 = - 2 \bar{\mathcal{D}}^{\alpha}
\mathcal{D}^{12}_{\alpha} \tilde{\Upsilon}_0 
= 
- 2 
\left[
 D^{12 \alpha} \Gamma^{(-)}_{0 \alpha} - \frac{1}{2}
\bar{\mathbb{D}}^{\alpha} \Gamma^{(-)}_{-1 \alpha} - \frac{1}{2} 
\{
\Gamma^{(-) \alpha}_{-1}, \Gamma^{(-)}_{0 \alpha}
\}
\right] 
\tilde{\Upsilon}_0,
\\
 & \bar{\tilde{\Upsilon}}_1 \overleftarrow{\mathcal{D}}^2 = 
2 \bar{\tilde{\Upsilon}}_0 
\overleftarrow{\mathcal{D}}^{12}_{\alpha}
\overleftarrow{\mathcal{D}}^{\alpha}
= 
2 \bar{\tilde{\Upsilon}}_0
 \left[
\frac{1}{2} 
\mathbb{D}^{\alpha}
\Gamma^{(-)}_{-1 \alpha}
+ 
D^{12 \alpha}
\Gamma^{(-)}_{-2 \alpha} 
+ \frac{1}{2}
\{
\Gamma^{(-) \alpha}_{-2}, \Gamma^{(-)}_{-1 \alpha}
\}
\right].
\end{aligned}
\end{eqnarray}
As we will see, the adjoint chiral superfields in the gauge multiplet
appear in these relations. We define the gauge covariantized chiral superfields
$\tilde{\Phi}_0, \bar{\tilde{\Phi}}_0$ in the gauge multiplet as 
\begin{eqnarray}
\tilde{\Phi}_0 &\equiv& 
- 2 
\left[
 D^{12 \alpha} \Gamma^{(-)}_{0 \alpha} - \frac{1}{2}
\bar{\mathbb{D}}^{\alpha} \Gamma^{(-)}_{-1 \alpha} - \frac{1}{2} 
\{
\Gamma^{(-) \alpha}_{-1}, \Gamma^{(-)}_{0 \alpha}
\}
\right]
\label{eq:N3_chiral}
, \\
\bar{\tilde{\Phi}}_0 &\equiv& 
2 \left[
\frac{1}{2} \mathbb{D}^{\alpha} \Gamma^{(-)}_{-1 \alpha}
+ D^{12 \alpha} \Gamma^{(-)}_{-2 \alpha} 
+ \frac{1}{2}
\{
\Gamma^{(-) \alpha}_{-2}, \Gamma^{(-)}_{-1 \alpha}
\}
\right].
\label{eq:N3_anti-chiral}
\end{eqnarray}
The gauge transformations of the fields $\tilde{\Phi}_0, 
\bar{\tilde{\Phi}}_0$ 
follow from the definition \eqref{eq:N3_chiral},
\eqref{eq:N3_anti-chiral}.
They are found to be 
\begin{eqnarray}
\tilde{\Phi}_0 \to e^{- i \bar{\lambda}_0} \tilde{\Phi}_0 e^{i
 \bar{\lambda}_0}, \qquad 
 \bar{\tilde{\Phi}}_0 \to e^{- i \bar{\lambda}_0} \bar{\tilde{\Phi}}_0
 e^{i \bar{\lambda}_0}.
\end{eqnarray}
As we will see, these gauge transformations are consistent with the fact
that the original (before gauge covariantized) field $\Phi_0, \bar{\Phi}_0$
transform as the adjoint representation of $G$.

The explicit form of the gauge covariantized 
chiral superfields can be obtained by the
perturbation in $V_{-1}, V_0, V_1$. 
Here we explicitly write down the connections up to $\mathcal{O} (V^3)$,
\begin{eqnarray}
\Gamma^{(-)}_{0 \alpha} 
&=& 
\bar{\mathbb{D}}_{\alpha} V_0 + \frac{1}{2} 
[V_0, \bar{\mathbb{D}}_{\alpha} V_0]
- \frac{1}{2} [V_{-1}, \bar{\mathbb{D}}_{\alpha} V_1]
+ \mathcal{O} (V^3), 
\label{eq:G0_3}
\\
\Gamma^{(-)}_{-1 \alpha}
&=& 
\mathbb{D}_{\alpha} V_{-1} 
 + \frac{1}{2} 
[V_0, \mathbb{D}_{\alpha} V_{-1}] 
+ \frac{1}{2}
[\mathbb{D}_{\alpha} V_0, V_{-1}]
+ \mathcal{O} (V^3), 
\label{eq:G1_3}
\\
\Gamma^{(-)}_{-2 \alpha} 
&=& 0.
\label{eq:G2_3}
\end{eqnarray}
Higher order corrections can be calculated systematically by using the
decomposition of $e^{\mathcal{V}^{[0]}}$ found in Appendix C.
Then the anti-chiral field is calculated as 
\begin{eqnarray}
\bar{\tilde{\Phi}}_0 = - \mathbb{D}^2 V_{-1} - \frac{1}{2} [V_0, \mathbb{D}^2 V_{-1}]
- \frac{1}{2} [\mathbb{D}^2 V_0, V_{-1}]
 - \{\mathbb{D}^{\alpha} V_0, \mathbb{D}_{\alpha}
 V_{-1}\}
+ \mathcal{O} (V^3),
\end{eqnarray}
while the chiral superfield $\tilde{\Phi}_0$ is 
\begin{eqnarray}
\tilde{\Phi}_0 &=& 
\bar{\mathbb{D}}^2 V_1 + \frac{1}{2} [V_0, \bar{\mathbb{D}}^2 V_1]
+ \{\bar{\mathbb{D}}^{\alpha} V_1, \bar{\mathbb{D}}_{\alpha} V_0 \}
+ \mathcal{O} (V^3).
\end{eqnarray}
Here we have used the constraints $\mathbb{D}_{\alpha} V_1 =
\bar{\mathbb{D}}_{\alpha} V_{-1} = 0$ on the tropical multiplet.
We stress that the anti-chiral superfield is written as the 
``$\mathbb{D}^2$-exact form'' $\bar{\tilde{\Phi}}_0 = \mathbb{D}^2 \hat{V}_{-1}$
and the gauge covariant chirality follows
from the nilpotency of the supercovariant derivative
$\mathbb{D}_{\alpha}$
for the full order in $V_{-1}, V_0, V_1$:
\begin{equation}
\mathcal{D}_{\alpha} \bar{\tilde{\Phi}}_0 
= \mathbb{D}_{\alpha} \mathbb{D}^2 \hat{V}_{-1} = 0.
\end{equation}
On the other hand, the chirality of $\tilde{\Phi}_0$ is shown 
in the perturbation of $V_{-1}, V_0, V_1$. 
For example, up to $\mathcal{O}(V^3)$, we can show that 
the superfield $\tilde{\Phi}_0$ satisfies the gauge covariantized
chirality condition,
\begin{eqnarray}
\bar{\mathcal{D}}_{\alpha} \tilde{\Phi}_0
&=& \bar{\mathbb{D}}_{\alpha} 
\left\{
\bar{\mathbb{D}}^2 V_1  
+ \frac{1}{2} [V_0, \bar{\mathbb{D}}^2 V_1]
+ \{\bar{\mathbb{D}}^{\beta} V_1,
\bar{\mathbb{D}}_{\beta} V_0 \}
\right\}
- [\bar{\mathbb{D}}_{\alpha} V_0, \bar{\mathbb{D}}^2 V_1]
+ \mathcal{O} (V^3)
\nonumber \\
&=& 0 + \mathcal{O} (V^3),
\end{eqnarray}
where we have used the projective constraints on $V_{-1}, V_0, V_1$.

The action of a hypermultiplet in the fundamental representation of the
gauge group $G$ is given in \eqref{eq:fundamental_hypermultiplet}. 
After expanding the fields in $\zeta$ and integrate over the 
$\mathbb{C}P^1$, we have the action in the $\mathcal{N} = 2$
subsuperspace as 
\begin{eqnarray}
S &=& \int \! d^3 x d^4 \theta 
\left[
\bar{\tilde{\Upsilon}}_0 \tilde{\Upsilon}_0 - \bar{\tilde{\Upsilon}}_1
\tilde{\Upsilon}_1 + \sum_{l=2}^{\infty} (-1)^l \bar{\tilde{\Upsilon}}_l
\tilde{\Upsilon}_l   
+ (\bar{\tilde{\Upsilon}}_1 \overleftarrow{\mathcal{D}}^2 -
\bar{\tilde{\Upsilon}}_0 \bar{\tilde{\Phi}}_0 ) \tilde{Y}_0
+ \bar{\tilde{Y}}_0 (\overrightarrow{\bar{\mathcal{D}}}^2 
\tilde{\Upsilon}_1 - \tilde{\Phi}_0 \tilde{\Upsilon}_0)
\right], \nonumber \\
\label{eq:N3_hyper_action}
\end{eqnarray}
where we have introduced the $\mathcal{N} = 2$ 
Lagrange multiplier superfields $Y_0, \tilde{Y}_0$ to impose the
projective constraints.
The gauge transformations of the Lagrange multipliers are defined as 
\begin{eqnarray}
\tilde{Y}_0 \to e^{- i \bar{\lambda}_0} \tilde{Y}_0, \quad 
 \bar{\tilde{Y}}_0 \to \bar{\tilde{Y}}_0 e^{ i \bar{\lambda}_0}.
\end{eqnarray}
The $\mathcal{N} = 2$ superfields 
$Y_0, \bar{\Phi}_0$ and $\bar{Y}_0, \Phi_0$ are 
interpreted as $\zeta^0$ components of 
the arctic and antarctic multiplets 
$Y, \bar{\Phi}$ and $\bar{Y}, \Phi$ respectively.
From the ``fields with tilde'', we can go back to the original 
$\mathcal{N} = 3$ projective superfields (before gauge covariantized) by factoring out
the decomposed fields $\hat{V}_{-}, \hat{V}_0, \hat{V}_{+}$ of the tropical multiplets:
\begin{eqnarray}
\begin{aligned}
 & \tilde{Y} = e^{\hat{V}_0} e^{\hat{V}_{+}} Y, \qquad 
 \bar{\tilde{Y}} = \bar{Y} e^{\hat{V}_{-}}, \\
 & \tilde{\Phi} = 
e^{\hat{V}_0} e^{\hat{V}_{+}} \Phi e^{- \hat{V}_{+}}
 e^{- \hat{V}_0}, \qquad 
 \bar{\tilde{\Phi}} = 
e^{- \hat{V}_{-}} \bar{\Phi} e^{\hat{V}_{-}}.
\end{aligned}
\end{eqnarray}
Especially, the $\zeta^0$ component of each field is found to be 
\begin{eqnarray}
\tilde{Y}_0 = e^{\hat{V}_0} Y_0, \quad \bar{\tilde{Y}}_0 =
 \bar{Y}_0, \quad 
\tilde{\Phi}_0 = 
 e^{\hat{V}_0} \Phi_0 e^{- \hat{V}_0}, 
\quad \bar{\tilde{\Phi}}_0 =
\bar{\Phi}_0.
\end{eqnarray}
The gauge transformations of the original fields are 
\begin{eqnarray}
\begin{aligned}
 & Y \to e^{- i \Lambda^{[0]}} Y, \quad \bar{Y} \to \bar{Y} e^{i
 \bar{\Lambda}^{[0]}}, \\
 & 
\Phi \to e^{- i \Lambda^{[0]}} \Phi e^{i \Lambda^{[0]}},
 \quad 
\bar{\Phi} \to 
e^{- i \bar{\Lambda}^{[0]}} \bar{\Phi} e^{i \bar{\Lambda}^{[0]}}.
\end{aligned}
\end{eqnarray}
As we have mentioned, $\Phi, \bar{\Phi}$ are adjoint representations of the gauge
group $G$. Note that $\Phi_0, \bar{\Phi}_0$ satisfy the ordinary (anti)chiral
superfield conditions $\bar{\mathbb{D}}_{\alpha} \Phi_0 =
\mathbb{D}_{\alpha} \bar{\Phi}_0 = 0$.

Now we rewrite the action \eqref{eq:N3_hyper_action} as follows.
First, we integrate out the infinite number of the auxiliary fields
$\tilde{\Upsilon}_l, \bar{\tilde{\Upsilon}}_{l}, \ (l \ge 2)$.
We then integrate out $\bar{\tilde{\Upsilon}}_1,
\tilde{\Upsilon}_1$ and dualize the fields $\tilde{\Upsilon}_1,
\bar{\tilde{\Upsilon}}_1$ into $\tilde{Y}_0, \bar{\tilde{Y}}_0$.
The action becomes 
\begin{eqnarray}
S &=& \int \! d^3 x d^4 \theta \ 
\left[
\bar{\tilde{\Upsilon}}_0 \tilde{\Upsilon}_0 + 
\overrightarrow{\bar{\mathcal{D}}}^2 \bar{\tilde{Y}}_0 \tilde{Y}_0
\overleftarrow{\mathcal{D}}^2 
- \bar{\tilde{\Upsilon}}_0 \bar{\tilde{\Phi}}_0 \tilde{Y}_0 - \bar{\tilde{Y}}_0 \Phi_0 \tilde{\Upsilon}_0
\right] \nonumber \\
& & \qquad - \frac{1}{4} \int \! d^3 x d^2 \theta \
\overrightarrow{\bar{\mathcal{D}}}^2 
\left[
- \bar{\tilde{Y}}_0 \Phi_0 \tilde{\Upsilon}_0
\right] 
- \frac{1}{4} \int \! d^3 x d^2 \bar{\theta} \
\left[
- \bar{\Phi}_0 \bar{\tilde{\Upsilon}}_0 \tilde{Y}_0
\right] 
\overleftarrow{\mathcal{D}}^2
\nonumber \\
&=& \int \! d^3 x d^4 \theta \ 
\left[
\bar{\Upsilon}_0 e^{\hat{V}_0} \Upsilon_0 + 
 \overrightarrow{\bar{\mathcal{D}}}^2
\bar{\tilde{Y}}_0 \cdot \tilde{Y}_0 \overleftarrow{\mathcal{D}}^2
\right] 
\nonumber \\
& & \qquad - \frac{1}{4} \int \! d^3 x d^2 \theta \
\left[
- \overrightarrow{\bar{\mathcal{D}}}^2 \bar{\tilde{Y}}_0
\tilde{\Phi}_0 \tilde{\Upsilon}_0
\right] 
- \frac{1}{4} \int \! d^3 x d^2 \bar{\theta} \
\left[
- \bar{\tilde{\Upsilon}}_0 
\bar{\tilde{\Phi}}_0
 \tilde{Y}_0 \overleftarrow{\mathcal{D}}^2 
\right].
\label{eq:N3hyper_rewrite}
\end{eqnarray}
where we have used the relations 
$\bar{\tilde{\Upsilon}}_0 = \bar{\Upsilon}_0$,
$\tilde{\Upsilon}_0 = e^{\hat{V}_0} \Upsilon_0$, 
the gauge covariant constraints $\mathcal{D}_{\alpha} \bar{\tilde{\Upsilon}}_0 = 
\bar{\mathcal{D}}_{\alpha} \tilde{\Upsilon}_0 = 0$ and 
the gauge covariant chiralities  
$\bar{\tilde{\Phi}} \overleftarrow{\mathcal{D}}_{\alpha}  = 
\overrightarrow{\bar{\mathcal{D}}}_{\alpha} \tilde{\Phi}
 = 0$. 
We stress that the covariant chirality of $\tilde{\Phi}_0,
\bar{\tilde{\Phi}}_0$ is crucial to write down the action in terms of  
the $\mathcal{N} = 2$ component superfields.
After the field redefinition 
\begin{eqnarray}
\overrightarrow{\bar{\mathcal{D}}}^2 \bar{\tilde{Y}}_0 \equiv T e^{-
 \hat{V}_0}, \quad 
\tilde{Y}_0 \overleftarrow{\mathcal{D}}^2 \equiv \bar{T}, 
\end{eqnarray}
the relabeling $\Upsilon_0 \to S$ and the rescaling $\Phi \to 8 \Phi$, 
we find that the action \eqref{eq:N3hyper_rewrite} precisely 
reproduces the action of 
the $\mathcal{N} = 3$ hypermultiplet coupled with non-Abelian gauge
multiplet \eqref{eq:N3hyper_N2superspace}.
Generalizations to the multi-flavour models, adjoint or bi-fundamental 
representations of hypermultiplets are straightforward.

\subsection{$\mathcal{N} = 4$ in three dimensions}
In this subsection, we 
generalize the $\mathcal{N} = 3$ construction to $\mathcal{N} =
4$ theories.
Since the left and right sectors are essentially the same, we concentrate on 
the left part.
In the Lindstr\"om-Ro\v{c}ek gauge, the weight-0 left tropical multiplet
$\mathcal{V}^{[0]}_L$ is expanded as 
\begin{equation}
\mathcal{V}^{[0]}_L = \frac{1}{\zeta_L} V_{L,-1} + V_{L,0} + \zeta_L V_{L,1}.
\end{equation}
In the following, we omit the subscript $L$. 
The non-Abelian gauge transformation is defined by \eqref{eq:gauge_transf_tropical}.
As in the $\mathcal{N} = 3$ case, we consider the decomposition
\eqref{eq:decomposition}. 
From the constraint $D^{[1] \bar{k}}_{\alpha} \mathcal{V}^{[0]} = 0$
we have an identity
\begin{eqnarray}
 e^{- \hat{V}_-} (D^{[1] \bar{k}}_{\alpha} e^{\hat{V}_{-}}) 
 = e^{\hat{V}_0} e^{\hat{V}_{+}} D_{\alpha}^{[1] \bar{k}} (e^{- \hat{V}_{+}} e^{-
 \hat{V}_0}).
 \label{eq:N4ident}
\end{eqnarray}
The both sides in \eqref{eq:N4ident} are expanded as 
\begin{eqnarray}
\begin{aligned}
e^{- \hat{V}_-} (D^{[1] \bar{k}}_{\alpha} e^{\hat{V}_{-}}) 
&\equiv \zeta \Gamma^{(-) \bar{k}}_{-1 \alpha} + \zeta^0
 \Gamma^{(-) \bar{k}}_{0 \alpha} + \sum^{\infty}_{l=1} \zeta^{-l}
 \Gamma^{(-) \bar{k}}_{l \alpha}, \\
e^{\hat{V}_0} e^{\hat{V}_{+}} D_{\alpha}^{[1] \bar{k}} (e^{- \hat{V}_{+}} e^{-
 \hat{V}_0}) 
&\equiv \sum^{\infty}_{l=0} \zeta^l \Gamma^{(+) \bar{k}}_{l \alpha}.
\end{aligned}
\end{eqnarray}
Therefore we have the following relations among the components
\begin{eqnarray}
\begin{aligned}
 & \Gamma^{(+) \bar{k}}_{1 \alpha} = \Gamma^{(-)\bar{k}}_{-1 \alpha},
 \quad \Gamma^{(+) \bar{k}}_{0 \alpha} = \Gamma^{(-) \bar{k}}_{0
 \alpha}, \\
 & \Gamma^{(+) \bar{k}}_{l \ge 2 \ \alpha} = \Gamma^{(-) \bar{k}}_{l \ge
 1 \ \alpha} = 0.
\end{aligned}
\end{eqnarray}
Using the $\zeta$-expansion of the supercovariant derivative 
 $D^{[1] \bar{k}}_{\alpha} = D^{2 \bar{k}}_{\alpha} - \zeta
D^{1 \bar{k}}_{\alpha}$ in the left hand side of \eqref{eq:N4ident}, 
we find that the $\zeta$ component in \eqref{eq:N4ident} vanishes identically,
\begin{equation}
\Gamma^{(-) \bar{k}}_{-1 \alpha} = 0.
\end{equation}
Therefore the gauge connection contains only terms with the zeroth order of $\zeta$,
\begin{equation}
\Omega^{\bar{k}}_{\alpha} \equiv 
e^{\hat{V}_0} e^{\hat{V}_{+}} D_{\alpha}^{[1] \bar{k}} (e^{- \hat{V}_{+}} e^{-
 \hat{V}_0}) = \zeta^0 \Gamma^{(-)\bar{k}}_{0 \alpha}.
\end{equation}
The gauge covariantized supercovariant derivative is defined as 
\begin{equation}
\mathcal{D}^{[1] \bar{k}}_{\alpha} \equiv D^{[1] \bar{k}}_{\alpha} +
 \Omega^{\bar{k}}_{\alpha} = 
 \mathcal{D}^{2 \bar{k}}_{\alpha} -
 \zeta \mathcal{D}^{1 \bar{k}}_{\alpha}.
\end{equation}
Then each component in the gauge covariant derivative is found to be
\begin{equation}
\mathcal{D}^{2 \bar{k}}_{\alpha} = D^{2 \bar{k}}_{\alpha} + \Gamma^{(-)
 \bar{k}}_{0 \alpha}, \qquad 
\mathcal{D}^{1 \bar{k}}_{\alpha} = D^{1 \bar{k}}_{\alpha}.
\end{equation}
As in the $\mathcal{N} = 3$ case, 
we define the left and right gauge covariant derivatives,
\begin{eqnarray}
\begin{aligned}
 & \overrightarrow{\mathcal{D}}^{[1] \bar{k}}_{\alpha} * 
= \overrightarrow{D}^{[1] \bar{k}}_{\alpha} * 
+ 
e^{q \hat{V}_0} e^{q \hat{V}_{+}} \overrightarrow{D}^{[1]
\bar{k}}_{\alpha} (e^{- q \hat{V}_{+}} e^{- q \hat{V}_0})
 \triangleright *, \\
 & * \overleftarrow{\mathcal{D}}^{[1]\bar{k}}_{\alpha} 
= * \overleftarrow{D}^{[1]\bar{k}}_{\alpha} 
+ * \triangleleft (e^{- q \hat{V}_{-}} 
\overleftarrow{D}^{[1]\bar{k}}_{\alpha} ) e^{q \hat{V}_{-}}.
\end{aligned}
\end{eqnarray}
We now calculate the explicit form of the gauge connection.
Form the left hand side in \eqref{eq:N4ident}, we find 
\begin{eqnarray}
e^{- \hat{V}_{-}} (D^{[1] \bar{k}}_{\alpha} e^{\hat{V}_{-}}) 
= - D^{1 \bar{k}}_{\alpha} \hat{V}_{-1}.
\end{eqnarray}
where we have used the fact that terms with negative powers of $\zeta$
vanish.
Then, we find 
\begin{equation}
\Gamma^{(-) \bar{k}}_{0 \alpha} 
= - D^{1 \bar{k}}_{\alpha} \hat{V}_{-1}.
\end{equation}
Therefore the gauge covariant derivatives are found to be 
\begin{eqnarray}
 \mathcal{D}^{1 \bar{1}}_{\alpha} = \mathbb{D}_{\alpha}, \quad 
 \mathcal{D}^{1 \bar{2}}_{\alpha} = D^{1 \bar{2}}_{\alpha}, \quad 
 \mathcal{D}^{2 \bar{1}}_{\alpha} = D^{2 \bar{1}}_{\alpha} 
- \mathbb{D}_{\alpha} \hat{V}_{-1}, \quad 
 \mathcal{D}^{2 \bar{2}}_{\alpha} = - \bar{\mathbb{D}}_{\alpha} 
- D^{1 \bar{2}}_{\alpha} \hat{V}_{-1}.
\end{eqnarray}
Again, 
$\hat{V}_{-1}$ is calculated by the perturbation of the components $V_{-1},
V_0, V_1$ (see Appendix B).
The anti-commutation relations of the covariant derivatives are found in
Appendix C.

In order to find the explicit form of the (anti)chiral superfields
$\Phi_0, \bar{\Phi}_0$ in the gauge multiplet, 
let us consider weight-1 left (ant)arctic multiplets
$\bar{\Upsilon}^{[1]}$, $\Upsilon^{[1]}$ in the 
(anti)fundamental representation.
The gauge invariant weight-2 Lagrangian is given by 
\begin{eqnarray}
\mathcal{L}^{[2]} = \bar{\Upsilon}^{[1]} e^{\mathcal{V}^{[0]}} \Upsilon^{[1]}.
\end{eqnarray}
The gauge transformation and the definition of the new fields 
\eqref{eq:N3_tilde} are the same in the $\mathcal{N}= 3$ case.
The projective constraints on the new fields are gauge covariantized.
Using the anti-commutation relations of the gauge covariant derivatives and the
gauge covariantized projective constraints, we find the following relations,
\begin{eqnarray}
\begin{aligned}
& \bar{\mathcal{D}}^2 \tilde{\Upsilon}_1 = 
- \bar{\mathcal{D}}^{\alpha} \mathcal{D}^{1 \bar{2}}_{\alpha}
\tilde{\Upsilon}_0 
= D^{1 \bar{2} \alpha} \Gamma^{(-) \bar{2}}_{0 \alpha}
 \tilde{\Upsilon}_0, \\
& \bar{\tilde{\Upsilon}}_1 
\overleftarrow{\mathcal{D}}^2
=  \bar{\tilde{\Upsilon}}_1 
\overleftarrow{\mathcal{D}}^{2 \bar{1}}_{\alpha}
\overleftarrow{\mathcal{D}}^{\alpha}
= 
\mathbb{D}^{\alpha} \Gamma^{(-) \bar{1}}_{0 \alpha}
\bar{\tilde{\Upsilon}}_0,
\end{aligned}
\end{eqnarray}
where we have defined 
\begin{equation}
\mathcal{D}^{1 \bar{1}}_{\alpha} \equiv \mathcal{D}_{\alpha} =
 \mathbb{D}_{\alpha}, \quad 
\mathcal{D}^{2 \bar{2}}_{\alpha} \equiv \bar{\mathcal{D}}_{\alpha} = 
- \bar{\mathbb{D}}_{\alpha} - D^{1 \bar{2}}_{\alpha} \hat{V}_{-1}.
\end{equation}
From these relations, we define the 
gauge covariantized chiral superfields $\tilde{\Phi}_0, \bar{\tilde{\Phi}}_0$,
\begin{eqnarray}
 \tilde{\Phi}_0 \equiv D^{1 \bar{2} \alpha} \Gamma^{(-) \bar{2}}_{0
 \alpha}, \qquad 
 \bar{\tilde{\Phi}}_0 \equiv \mathbb{D}^{\alpha} \Gamma^{(-) \bar{1}}_{0
\alpha}.
\end{eqnarray}
Since the $\zeta^0$ component of the gauge connection 
is given by $\Gamma^{(-) \bar{1}}_{0 \alpha} = - \mathbb{D}_{\alpha}
\hat{V}_{-1}$, the anti-chiral superfield $\bar{\tilde{\Phi}}_0$ 
is again $\mathbb{D}^2$-exact form. Therefore 
the chirality $\mathcal{D}_{\alpha} \bar{\tilde{\Phi}}_0 = \mathbb{D}_{\alpha}
\bar{\tilde{\Phi}}_0 = 0$ is shown for the full order in $V_{-1}, V_0, V_1$.
On the other hand, the chirality of $\tilde{\Phi}_0$ is shown in the
perturbation of $V_{-1}, V_0, V_1$. 
For example, up to $\mathcal{O} (V^3)$, we have 
\begin{eqnarray}
\bar{\mathcal{D}}_{\alpha} \tilde{\Phi}_0 
&=& 
 \bar{\mathbb{D}}_{\alpha} D^{1 \bar{2} \beta} D^{1
\bar{2}}_{\beta} V_{-1} 
+
 \frac{1}{2} \bar{\mathbb{D}}_{\alpha} D^{1 \bar{2} \beta} D^{1
\bar{2}}_{\beta}
[V_0, V_{-1}] 
+
\left[
 D^{1 \bar{2}}_{\alpha} V_{-1},
 D^{1 \bar{2} \beta} D^{1
\bar{2}}_{\beta} V_{-1}
\right]
+ \mathcal{O} (V^3)
\nonumber \\
&=& 0 + \mathcal{O} (V^3),
\end{eqnarray}
where we have used the relations obtained by the repeated use of the
constraints \eqref{Okk_constraint}:
\begin{equation}
\bar{\mathbb{D}}^2 V_0 = - \bar{\mathbb{D}}^{\alpha} D^{1
 \bar{2}}_{\alpha} V_{-1} = 
D^{1 \bar{2}}_{\alpha} \bar{\mathbb{D}}^{\alpha} V_{-1} = 0.
\end{equation}
Now we have proved the chirality of $\bar{\tilde{\Phi}}_0$, $\tilde{\Phi}_0$.
Note that $\Phi_0, \bar{\Phi}_0$ obtained from 
the left tropical multiplet by this way belong to the right multiplet
\cite{KuLiTa, ArSa}. 
The construction of the action is the same in the $\mathcal{N} = 3$ case.

\subsection{$\mathcal{N} = 2$ in four dimensions}
The tropical multiplet 
$\mathcal{V} = \sum_{l=-\infty}^{\infty} \zeta^l V_l$ 
satisfy the following projective constraints,
\begin{eqnarray}
\nabla_{\alpha} \mathcal{V} = \bar{\nabla}_{\dot{\alpha}} \mathcal{V} =
 0.
\label{eq:4dN2_constraints}
\end{eqnarray}
We consider the decomposition 
\eqref{eq:decomposition} of the tropical multiplet.
Using the constraints \eqref{eq:4dN2_constraints} 
we have the following identities,
\begin{eqnarray}
& & e^{- \hat{V}_{-}} (\nabla_{\alpha} e^{\hat{V}_{-}}) = 
e^{\hat{V}_0} e^{\hat{V}_{+}} \nabla_{\alpha} (e^{- \hat{V}_{+}} e^{-
\hat{V}_{0}}), 
\label{eq:4d_ident1}
\\
& & e^{- \hat{V}_{-}} (\bar{\nabla}_{\dot{\alpha}} e^{\hat{V}_{-}})
= e^{\hat{V}_{0}} e^{\hat{V}_{+}} \bar{\nabla}_{\dot{\alpha}} (e^{-
\hat{V}_{+}} e^{- \hat{V}_{0}}).
\label{eq:4d_ident2}
\end{eqnarray}
The definition of the gauge connections is 
\begin{eqnarray}
\begin{aligned}
\Omega_{\alpha} \equiv& 
e^{q \hat{V}_0} e^{q \hat{V}_{+}} \nabla_{\alpha} 
(e^{- q \hat{V}_{+}} e^{- q \hat{V}_{0}} ) = e^{- q \hat{V}_{-}}
(\nabla_{\alpha} e^{q \hat{V}_{-}}), 
\\
\bar{\Omega}_{\dot{\alpha}} \equiv&
e^{q \hat{V}_0} e^{q \hat{V}_{+}} \bar{\nabla}_{\dot{\alpha}} (e^{- q
\hat{V}_{+}} e^{- q \hat{V}_0}) = e^{- q \hat{V}_{-}}
(\bar{\nabla}_{\dot{\alpha}} e^{q \hat{V}_{-}}).
\end{aligned}
\end{eqnarray}
By the same way, the gauge covariantized left supercovariant derivatives are defined as 
\begin{eqnarray}
\begin{aligned}
\overrightarrow{\nabla}^G_{\alpha} * \equiv& 
\nabla_{\alpha} * + \Omega_{\alpha} \triangleright * 
\equiv (\mathcal{D}_{1 \alpha} + \zeta \mathcal{D}_{2 \alpha}) *, \\
\overrightarrow{\bar{\nabla}}^G_{\dot{\alpha}} * \equiv& \bar{\nabla}_{\dot{\alpha}} * +
 \bar{\Omega}_{\dot{\alpha}} \triangleright * \equiv 
(\bar{\mathcal{D}}^2_{\dot{\alpha}} - \zeta
\bar{\mathcal{D}}^1_{\dot{\alpha}}) *.
\end{aligned}
\end{eqnarray}
The right derivatives are defined similarly.
From the middle and the left sides in \eqref{eq:4d_ident1} we have the
$\zeta$ expansions, 
\begin{eqnarray}
e^{- \hat{V}_{-}} (\nabla_{\alpha} e^{\hat{V}_{-}})
&\equiv&  
\Gamma^{(-)}_{-1 \alpha} \zeta + \Gamma^{(-)}_{0 \alpha} \zeta^0 + \sum^{\infty}_{l=1}
\Gamma^{(-)}_{l  \alpha} \zeta^{-l}, \\
e^{\hat{V}_0} e^{\hat{V}_{+}} \nabla_{\alpha} (e^{- \hat{V}_{+}} e^{-
\hat{V}_{0}})
&\equiv& 
\sum^{\infty}_{l=0} \Gamma^{(+)}_{l  \alpha} \zeta^l.
\end{eqnarray}
By the same way, we define $\bar{\Gamma}^{(\pm)}_{l \dot{\alpha}}$ 
by the $\zeta$ expansions of \eqref{eq:4d_const2}.
Then the following relations hold,
\begin{eqnarray}
\begin{aligned}
& \Gamma^{(-)}_{0 \alpha} = \Gamma^{(+)}_{0 \alpha}, \quad 
\Gamma^{(-)}_{l \ge 1 \ \alpha} = \Gamma^{(+)}_{l \ge 1 \ \alpha} =
 0, \\
& \bar{\Gamma}^{(-)}_{0 \dot{\alpha}} = \bar{\Gamma}^{(+)}_{0 \dot{\alpha}}, \quad 
\bar{\Gamma}^{(-)}_{l \ge 1 \ \dot{\alpha}} = \bar{\Gamma}^{(+)}_{l \ge
 1 \ \dot{\alpha}} =  0.
\end{aligned}
\end{eqnarray}
We can show that the $\zeta^1$ terms in the left hand side of
\eqref{eq:4d_ident1} and \eqref{eq:4d_ident2} vanish identically
$\Gamma^{(-)}_{-1 \alpha} = \bar{\Gamma}^{(-)}_{-1 \dot{\alpha}} = 0$.
Therefore the gauge covariantized supercovariant derivatives are found to be 
\begin{eqnarray}
\begin{array}{l}
\mathcal{D}_{1 \alpha} = D_{1 \alpha} + \Gamma^{(-)}_{0 \alpha}, \\
\mathcal{D}_{2 \alpha} = D_{2 \alpha},
\end{array}
\qquad 
\begin{array}{l}
\bar{\mathcal{D}}^2_{\dot{\alpha}} = \bar{D}^2_{\dot{\alpha}} +
\bar{\Gamma}^{(-)}_{0 \dot{\alpha}}, \\
\bar{\mathcal{D}}^1_{\dot{\alpha}} = \bar{D}^1_{\dot{\alpha}}.
\end{array}
\end{eqnarray}
The anti-commutation relations of the gauge covariant derivatives are
found in Appendix B.

The gauge connections are calculated as 
\begin{eqnarray}
\begin{aligned}
\Omega_{\alpha} =& e^{- \hat{V}_{-}} (\nabla_{\alpha} e^{\hat{V}_{-}})
= D_{2 \alpha} \hat{V}_{-1}, \\
\bar{\Omega}_{\dot{\alpha}}
=& e^{- \hat{V}_{-}} (\bar{\nabla}_{\dot{\alpha}} e^{\hat{V}_{-}})
= - \bar{D}^1_{\dot{\alpha}} \hat{V}_{-1}.
\end{aligned}
\end{eqnarray}
The explicit expressions of the connections $\Omega_{\alpha}, \bar{\Omega}_{\dot{\alpha}}$ 
depend only on the $\zeta^{-1}$ term $\hat{V}_{-1}$ in the
decomposition of the tropical multiplet. 
This is the same situation found in the three-dimensional case.
Up to $\mathcal{O} (V^3)$, we have 
\begin{eqnarray}
\begin{aligned}
 & 
\Gamma^{(-)}_{0 \alpha} = D_{2 \alpha} \hat{V}_{-1} = D_{2 \alpha}
\left(
V_{-1} + \frac{1}{2} [V_0, V_{-1}] 
\right)
+ \mathcal{O} (V^3), \\
 & 
\bar{\Gamma}^{(-)}_{0 \dot{\alpha}} = 
- \bar{D}^1_{\dot{\alpha}} \hat{V}_{-1} = 
- \bar{D}^1_{\dot{\alpha}} 
\left(
V_{-1} + \frac{1}{2} [V_0, V_{-1}]
\right)
+ \mathcal{O} (V^3).
\end{aligned}
\end{eqnarray}
As in the three-dimensional cases, 
we define the new fields from the (ant)arctic 
multiplets:
\begin{eqnarray}
 \tilde{\Upsilon} \equiv e^{\hat{V}_0} e^{\hat{V}_{+}} \Upsilon, \qquad 
 \bar{\tilde{\Upsilon}} \equiv \bar{\Upsilon} e^{\hat{V}_{-}}.
\end{eqnarray}
The first (the second) fields satisfy the left (right) gauge
covariantized projective constraint.

Using the gauge covariantized constraints on the component superfields  $\tilde{\Upsilon}_0,
\tilde{\Upsilon}_0$, $\tilde{\Upsilon}_1, \bar{\tilde{\Upsilon}}_1$
and the anti-commutation relations of the gauge covariant derivatives, 
we obtain the following relations,
\begin{eqnarray}
\begin{aligned}
 & (\bar{\mathcal{D}}^2)^2 \tilde{\Upsilon}_1 
= \bar{\mathcal{D}}^2_{\dot{\alpha}} \bar{\mathcal{D}}^{2 \dot{\alpha}}
\tilde{\Upsilon}_1
= \bar{\mathcal{D}}^2_{\dot{\alpha}} \bar{\mathcal{D}}^{1 \dot{\alpha}}
\tilde{\Upsilon}_0 
= -\bar{D}_{ \dot{\alpha}} \bar{\Gamma}^{(-) \dot{\alpha}}_{0} \tilde{\Upsilon}_0, \\
 & \bar{\tilde{\Upsilon}}_1 (\overleftarrow{\mathcal{D}}^2)^2
= \bar{\tilde{\Upsilon}}_1 
\overleftarrow{\mathcal{D}}_{\alpha 2}
\overleftarrow{\mathcal{D}}^{\alpha}_2
= \bar{\tilde{\Upsilon}}_0 
\overleftarrow{\mathcal{D}}_{\alpha 1}
\overleftarrow{\mathcal{D}}^{\alpha}_{2} 
= 
\bar{\tilde{\Upsilon}}_0
D_{2}^{\alpha} \Gamma^{(-)}_{0 \alpha}.
\end{aligned}
\end{eqnarray}
From these expressions, we define the gauge covariantized chiral superfields,
\begin{eqnarray}
\tilde{\Phi}_0 \equiv 
- \bar{D}^1_{\dot{\alpha}} \bar{\Gamma}^{(-) \dot{\alpha}}_0, 
\qquad 
\bar{\tilde{\Phi}}_0 \equiv 
D^{\alpha}_2 \Gamma^{(-)}_{0 \alpha}.
\end{eqnarray}
In the following, we define the supercovariant derivative in the 
$\mathcal{N} = 1$ subsuperspace as 
$\mathbb{D}_{\alpha} = D_{2 \alpha}, \bar{\mathbb{D}}_{\dot{\alpha}} =
\bar{D}^2_{\dot{\alpha}}$.
We also write the corresponding gauge covariantized supercovariant derivatives as 
$\mathcal{D}_{\alpha}, \bar{\mathcal{D}}_{\dot{\alpha}}$.
Then up to cubic order in $V_{-1}, V_0, V_1$, we obtain
\begin{eqnarray}
& & \tilde{\Phi}_0 = \bar{D}^1_{\dot{\alpha}} \bar{D}^{1 \dot{\alpha}} 
\left(
V_{-1} + \frac{1}{2} [V_0, V_{-1}]
\right)
+ \mathcal{O} (V^3), \\
& & \bar{\tilde{\Phi}}_0 = D^{\alpha}_2 D_{2 \alpha} \hat{V}_{-1}
= \mathbb{D}^2 \hat{V}_{-1}.
\end{eqnarray}
Therefore we find
\begin{eqnarray}
\mathcal{D}_{\alpha} \bar{\tilde{\Phi}}_0 = \mathbb{D}_{\alpha}
 \mathbb{D}^2 \hat{V}_{-1} = 0.
\end{eqnarray}
Again, the chirality of $\bar{\tilde{\Phi}}_0$ is shown in the full
order of $V_{-1}, V_0, V_{1}$.
On the other hand, the chirality of $\tilde{\Phi}_0$ is shown perturbatively,
\begin{eqnarray}
\bar{\mathcal{D}}_{\dot{\alpha}} \tilde{\Phi}_0
&=&
\bar{\mathbb{D}}_{\dot{\alpha}} \bar{D}^1_{\dot{\beta}} \bar{D}^{1
\dot{\beta}} V_{-1}
+
\frac{1}{2} \bar{\mathbb{D}}_{\dot{\alpha}}
\bar{D}^1_{\dot{\beta}} \bar{D}^{1 \dot{\beta}} [V_0, V_{-1}]
- [\bar{D}^1_{\dot{\alpha}} V_{-1}, \bar{D}^1_{\dot{\beta}} \bar{D}^{1
\dot{\beta}} V_{-1}] 
+ \mathcal{O} (V^3) \nonumber \\
&=& 0 + \mathcal{O} (V^3).
\end{eqnarray}
The action is constructed as in the cases of three dimensions.

\section{Conclusion and discussions}
In this paper we have studied the explicit relations between the
non-Abelian gauge multiplets and 
the tropical multiples in three-dimensional $\mathcal{N} = 3$,
$\mathcal{N} = 4$ and four-dimensional $\mathcal{N} = 2$ projective superspaces.
In three dimensions, the actions possess the conformal invariance.
Although the formal procedure to construct the gauge invariant actions
has been discussed in the literature, the explicit form of the
decomposition of the non-Abelian tropical multiplet $e^{\mathcal{V}} =
e^{\hat{V}_{-}} e^{\hat{V}_0} e^{\hat{V}_{+}}$ and the 
expression of the gauge connection $\Omega_{\alpha}$ have not been studied in detail. 
For Abelian gauge groups, the decomposition of the tropical multiplet is
trivial, namely, in the Lindstr\"om-Ro\v{c}ek gauge it is given by 
$\hat{V}_{-} = V_{-1} \zeta^{-1}$, $\hat{V}_0 = V_0$,
$\hat{V}_{+} = \zeta V_1$. 
Then the $\mathcal{N} = 2$ vector superfield $V$ is identified
with the $\zeta^0$ component 
 $V_0$ in the tropical multiplet. 
The relations \eqref{eq:abelian_relation} 
among the chiral superfields $\Phi_0, \bar{\Phi}_0$ in the gauge multiplet 
and $V_1, V_{-1}$ in the tropical multiplet are linear and the
chiralities of $\Phi_0, \bar{\Phi}_0$ follows automatically.

On the other hand, for non-Abelian gauge groups, 
the decomposition is quite non-trivial and the relation between the
gauge multiplet and the tropical multiplet becomes highly non-linear.
In the present paper, we have performed the decomposition of the
tropical multiplet for non-Abelian gauge groups explicitly 
and found the precise form of the 
components $\hat{V}_{-},
\hat{V}_0, \hat{V}_{+}$.
Using the decomposition, the gauge connection 
$\Omega_{\alpha}$ has been constructed.
We have then considered 
a hypermultiplet coupled to the gauge multiplet.
The gauge covariantized projective constraints on the (ant)arctic
multiplets and the algebras of the gauge covariantized 
supercovariant derivatives enable us to define 
the adjoint superfields $\tilde{\Phi}_0, \bar{\tilde{\Phi}}_0$ 
in the gauge multiplet.
We have written down the expressions of $\tilde{\Phi}_0,
\bar{\tilde{\Phi}}_0$ in terms of the component fields in the tropical
multiplet. 
The gauge covariantized chiral conditions of $\tilde{\Phi}_0,
\bar{\tilde{\Phi}}_0$ have been shown. 
The chirality of $\bar{\tilde{\Phi}}_0$ is shown to be holds for the full order in
$V_{-1}, V_0, V_1$ 
while that of $\tilde{\Phi}_0$ should be proved perturbatively. 
We have demonstrated that the chirality of $\tilde{\Phi}_0$ 
holds up to $\mathcal{O} (V^3)$
in the non-Abelian gauge group.
However, the higher order calculations are possible by using the 
explicit decompositions presented in Appendix B.

Compare to the Abelian gauge group, the calculation is quite non-linear
and needs precise treatment. 
As we have shown in the Appendix B, the explicit form of the decomposition
$e^{\mathcal{V}} = e^{\hat{V}_{-}} e^{\hat{V}_0} e^{\hat{V}_{+}}$ is
obtained iteratively. The gauge connections depend only on the $\zeta,
\zeta^2$ ($\zeta$) components of $\hat{V}_{-}$ 
in $\mathcal{N} = 3$ ($\mathcal{N} = 4$)
in three dimensions. 
Similarly, the adjoint superfields $\tilde{\Phi}_0$,
$\bar{\tilde{\Phi}}_0$ are completely determined by $\hat{V}_{-1},
\hat{V}_{-2}$ ($\hat{V}_{-1}$) in $\mathcal{N} = 3$ ($\mathcal{N} = 4$).
We have also calculated the gauge connections in the four-dimensional
$\mathcal{N} = 2$ projective superspace. 
Chiralities of the adjoint superfields in the gauge multiplet have been
shown also in four dimensions.

We believe our study provides useful insights into the future researches of gauge
theories in the projective superspace formalism. 
For example, non-Abelian superconformal Chern-Simons theories in
three-dimensions are an interesting topic.
Generalizations to higher dimensional theories such as five and six
dimensions cases \cite{KuLi, Ku2} are also interesting.
We will come back to these issues in near future.

\subsection*{Acknowledgments}
The work of M.~A is supported in part by the Research Program
MSM6840770029, by the project of International Cooperation ATLAS-CERN of the Ministry
of Education, Youth and Sports of the Czech Republic, and by the Japan Society for the
Promotion of Science (JSPS) and Academy of Sciences of the Czech Republic (ASCR) under the Japan -
Czech Republic Research Cooperative Program.


\begin{appendix}
\section{
Conventions and notations of ordinary superspaces
}\label{appendixA}
\subsection{Three dimensions}
In this appendix, we present the basic conventions 
and notations of the standard $\mathcal{N} =2$, $\mathcal{N} = 3$ and 
$\mathcal{N} = 4$ superspaces in three dimensions.
We use the mostly plus convention of the three dimensional metric $\eta_{mn} = 
\mathrm{diag} (-1,+1,+1)$.
The three-dimensional $\mathcal{N} = 2$ superspace is represented by the 
coordinates $z^A=(x^{m}, \theta^{\alpha}, \bar{\theta}^{\alpha})$ where 
$\theta, \bar{\theta}$ are two component spinors. 
The spinor indices are raised and lowered by the anti-symmetric 
symbol $\varepsilon^{12} = - \varepsilon_{12} = 1$. 
The gamma matrices are defined by 
$(\gamma^{m})_{\alpha} {}^{\beta} = (i \tau^2, \tau^1 ,\tau^3)$ 
which satisfies the Clifford algebra 
$\{\gamma^{m}, \gamma^{n} \} = 2 \eta^{mn}$.
Here $\tau^I \ (I=1,2,3)$ are the Pauli matrices.
The supercovariant derivatives in $\mathcal{N} = 3$ superspace are defined by 
\begin{eqnarray}
\begin{aligned}
& D^{ij}_{\alpha} = \frac{\partial}{\partial \theta^{\alpha}_{ij}} 
+ i \theta^{\beta}_{ij} \partial_{\alpha \beta},\qquad 
\partial_{\alpha\beta}\equiv \gamma^m_{\alpha\beta}\partial_m, \\
& \{D^{ij}_{\alpha}, D^{kl}_{\beta} \} = - 2 i \varepsilon^{i(k}
 \varepsilon^{l)j} \partial_{\alpha \beta}.
\end{aligned}
\end{eqnarray}
The supercovariant derivatives in $\mathcal{N} = 4$ superspace are
defined by 
\begin{eqnarray}
\begin{aligned}
& D^{i \bar{j}}_{\alpha} = \frac{\partial}{\partial \theta^{\alpha}_{i
 \bar{j}}} + i \theta^{\beta}_{i \bar{j}} \partial_{\alpha \beta}, \\
& \{D^{i \bar{j}}_{\alpha}, D^{k \bar{l}}_{\beta} \} = 
 2 i \varepsilon^{ik} \varepsilon^{\bar{j} \bar{l}} \partial_{\alpha
 \beta}.
\end{aligned}
\end{eqnarray}
The supercovariant derivatives in $\mathcal{N} = 2$ superspace 
are defined by 
\begin{eqnarray}
\begin{aligned}
& \mathbb{D}_{\alpha} = \partial_{\alpha} + i (\gamma^{m} \bar{\theta})_{\alpha} 
\partial_{m}, \qquad 
\bar{\mathbb{D}}_{\alpha} = - \bar{\partial}_{\alpha} - i (\theta 
\gamma^{m})_{\alpha} \partial_{m}, \label{cov1} \\
& \{\mathbb{D}_{\alpha}, \bar{\mathbb{D}}_{\beta} \} = - 2 i
 \gamma^{m}_{\alpha \beta} \partial_{m}, \quad \{ \mathbb{D}_{\alpha},
 \mathbb{D}_{\beta} \} = \{\bar{\mathbb{D}}_{\alpha},
 \bar{\mathbb{D}}_{\beta} \} = 0.
\end{aligned}
\end{eqnarray}
The Grassmann measure of integration in the $\mathcal{N} = 2$ superspace
is defined by 
\begin{eqnarray}
d^2 \theta = - \frac{1}{4} d \theta^{\alpha} d \theta_{\alpha}, \quad 
d^2 \bar{\theta} = - \frac{1}{4} d \bar{\theta}^{\alpha} d 
\bar{\theta}_{\alpha}, \quad d^4 \theta = d^2 \theta d^2 \bar{\theta}.
\end{eqnarray}
They are normalized such that,
\begin{eqnarray}
\int \! d^2 \theta \ \theta^2 = 1, \quad \int \! d^2 \bar{\theta} \ 
\bar{\theta}^2=1, \quad \int \! d^4 \theta \ \theta^2 \bar{\theta}^2 = 1.
\end{eqnarray}
Within the space-time integration, the following relation holds,
\begin{eqnarray}
\int \! d^4 \theta \ F(z) = \left. \frac{1}{16} (\mathbb{D}^2 \bar{\mathbb{D}}^2 F(z)) 
\right|_{\theta = \bar{\theta} = 0},
\end{eqnarray}
where $F(z)$ is an $\mathcal{N} = 2$ superfield. 
The chiral and anti-chiral coordinates are defined by 
\begin{eqnarray}
x^{m}_L = x^{m} + i \theta \gamma^{m} \bar{\theta}, \qquad 
x^{m}_R = x^{m} - i \theta \gamma^{m} \bar{\theta}.
\end{eqnarray}
\end{appendix}
We use the following relations among the $\mathcal{N} = 2$, 
$\mathcal{N} = 3$ and $\mathcal{N} = 4$ superspaces \cite{KuPaTaUn}:
\begin{eqnarray}
& & \theta^{\alpha} = \theta^{\alpha}_{11} = \theta^{\alpha}_{1\bar{1}},
 \quad 
\bar{\theta}^{\alpha} = \theta^{\alpha}_{22} =
\theta^{\alpha}_{2\bar{2}}, \\
& & \mathbb{D}_{\alpha} = D^{11}_{\alpha} = D^{1 \bar{1}}_{\alpha},
 \quad 
\bar{\mathbb{D}}_{\alpha} = - D^{22}_{\alpha} = - D^{2\bar{2}}_{\alpha}.
\end{eqnarray}

\subsection{Four dimensions}
In four dimensions, we use the metric 
$\eta_{mn} = \mathrm{diag} (-1,1,1,1)$. 
We follow the Wess-Bagger convention \cite{WeBa} in the $\mathcal{N} = 1$ superspace.
The supercovariant derivative is defined by 
\begin{eqnarray}
D_{\alpha} &=& \frac{\partial}{\partial \theta^{\alpha}} 
+ i  \sigma^m_{\alpha \dot{\alpha}} \partial_m, \\
\bar{D}_{\dot{\alpha}} &=& - \frac{\partial}{\partial \bar{\theta}^{\dot{\alpha}}} 
 - i \theta^{\alpha} \sigma^m_{\alpha \dot{\alpha}} \partial_m.
\end{eqnarray}
They satisfy the following algebra,
\begin{eqnarray}
& & \{ D_{\alpha}, \bar{D}_{\dot{\alpha}} \} = - 2 i \sigma^m_{\alpha
 \dot{\alpha}} \partial_m, \\
& & \{D_{\alpha}, D_{\beta} \} = \{\bar{D}_{\dot{\alpha}},
 \bar{D}_{\dot{\beta}} \} = 0.
\end{eqnarray}

\section{Decomposition of tropical multiplet}
In this appendix, we calculate the decomposition of the tropical multiplet,
\begin{equation}
e^{\mathcal{V}^{[0]}} = e^{\hat{V}_{-}} e^{\hat{V}_0} e^{\hat{V}_{+}}.
\end{equation}
In the Lindstr\"om-Ro\v{c}ek gauge, we have the following expansion in $\zeta$,  
\begin{eqnarray}
& & \mathcal{V}^{[0]} = \mathcal{V}_{-} + \mathcal{V}_0 + \mathcal{V}_{+}, \\
& & \mathcal{V}_{+} \equiv\zeta V_1, \quad 
\mathcal{V}_0 \equiv V_0, \quad 
\mathcal{V}_{-} \equiv \frac{1}{\zeta} V_{-1}.
\end{eqnarray}
The calculation is performed in the perturbation of the components in
the tropical multiplet $V_{-1}, V_0, V_{-1}$.
At leading (Abelian) order, we have $\hat{V}_{-} =
\mathcal{V}_{-}$, $\hat{V}_0 = \mathcal{V}_0$, $\hat{V}_{+} =
\mathcal{V}_{+}$. 
In the following we will determine $\hat{V}_{\pm},
\hat{V}_0$ at the quadratic and cubic orders in $V_{-1}, V_0, V_{1}$.
\\
\\
\underline{\textbullet $\mathcal{O}(V^2)$ calculation}
\\
\\
We determine the functions $f^{(2)}_0, f^{(2)}_{\pm}$ 
that satisfy the following relation up to terms in $\mathcal{O}(V^3)$,
\begin{equation}
e^{\mathcal{V}^{[0]}} = e^{\mathcal{V}_{-} + f^{(2)}_{-}}
 e^{\mathcal{V}_0 + f^{(2)}_0} e^{\mathcal{V}_{+} + f^{(2)}_{+}}.
\end{equation}
The both sides in the above are calculated as 
\begin{eqnarray}
& & e^{\mathcal{V}^{[0]}} = 1 + \mathcal{V}_{-} + \mathcal{V}_0 + \mathcal{V}_{+} 
\nonumber \\
& & \qquad + \frac{1}{2} 
\left(
\mathcal{V}_{+}^2 + \mathcal{V}_0^2 + \mathcal{V}_{-}^2 + 
\mathcal{V}_{+} \mathcal{V}_{0} + \mathcal{V}_0 \mathcal{V}_{+} +
\mathcal{V}_0 \mathcal{V}_{-} + \mathcal{V}_{-} \mathcal{V}_{0} +
\mathcal{V}_{-} \mathcal{V}_{+} + \mathcal{V}_{+} \mathcal{V}_{-}
\right)
+ \mathcal{O} (\mathcal{V}^3), \\
& & e^{\mathcal{V}_{-} + f^{(2)}_{-}}
 e^{\mathcal{V}_0 + f^{(2)}_0} e^{\mathcal{V}_{+} + f^{(2)}_{+}} 
\nonumber \\
& & 
\qquad 
= 
1 + \mathcal{V}_{-} + \mathcal{V}_0 + \mathcal{V}_{+} \nonumber \\
& & \qquad + 
\left(
\mathcal{V}_0 \mathcal{V}_+ + \frac{1}{2} \mathcal{V}_{+}^2 +
f^{(2)}_{+}
\right)
+
\left(
\mathcal{V}_{-} \mathcal{V}_{+} + \frac{1}{2} \mathcal{V}_0^2 + f^{(2)}_0
\right)
+
\left(
\mathcal{V}_{-} \mathcal{V}_0 + \frac{1}{2} \mathcal{V}_{-}^2 
+ f^{(2)}_{-}
\right)
+ \mathcal{O} (\mathcal{V}^3).
\nonumber \\
\end{eqnarray}
Note that all the terms are noncommutative and the ordering of the
product is important.
Comparing these results, we find the following expressions of the functions,
\begin{eqnarray}
f^{(2)}_{+} = 
\frac{1}{2} \zeta [V_1, V_0], \qquad 
f^{(2)}_0 = \frac{1}{2} [V_1, V_{-1}], \qquad 
f^{(2)}_{-} = \frac{1}{2} \zeta^{-1} [V_0, V_{-1}].
\label{eq:O2sol}
\end{eqnarray}
\\
\\
\underline{\textbullet $\mathcal{O}(V^3)$ calculation}
\\
\\
Next, we determine the functions $f^{(3)}_{\pm}, f^{(3)}_0$ that satisfy
the following relation up to $\mathcal{O}(V^4)$, 
\begin{equation}
e^{\mathcal{V}^{[0]}} = e^{\mathcal{V}_{-} + f^{(2)}_{-} + f^{(3)}_{-}}
 e^{\mathcal{V}_0 + f^{(2)}_0 + f^{(3)}_0} e^{\mathcal{V}_{+} +
 f^{(2)}_{+} + f^{(3)}_{+}},
\label{eq:cubic}
\end{equation}
where $f^{(2)}_{\pm}, f^{(2)}_0$ are functions \eqref{eq:O2sol} at $\mathcal{O} (V^2)$.
Therefore at the order $\mathcal{O}(\mathcal{V}^2)$ the above relation holds automatically.
At $\mathcal{O} (V^3)$, we find the left hand side in \eqref{eq:cubic} is 
\begin{eqnarray}
& & e^{\mathcal{V}^{[0]}}|_{\mathcal{O}(\mathcal{V}^3)} 
\nonumber \\
& & = 
\frac{1}{3!} 
\left(
  \mathcal{V}_{+}^3 
+ \mathcal{V}_0^2 \mathcal{V}_{+} 
+ \mathcal{V}_{+} \mathcal{V}_0 \mathcal{V}_{+} 
+ \mathcal{V}_0 \mathcal{V}_{+}^2
+ \mathcal{V}_{-} \mathcal{V}_{+}^2 
+ \mathcal{V}_{+} \mathcal{V}_{-} \mathcal{V}_{+}
+ \mathcal{V}_{+}^2 \mathcal{V}_{0}
+ \mathcal{V}_{+} \mathcal{V}_0^2 
+ \mathcal{V}_0 \mathcal{V}_{+} \mathcal{V}_0 
+ \mathcal{V}_{+}^2 \mathcal{V}_{-} 
\right)
\nonumber \\
& & 
+ \frac{1}{3!} 
\left(
  \mathcal{V}_0^3 
+ \mathcal{V}_0 \mathcal{V}_{-} \mathcal{V}_{+}
+ \mathcal{V}_{-} \mathcal{V}_0 \mathcal{V}_{+}
+ \mathcal{V}_{-} \mathcal{V}_{+} \mathcal{V}_0 
+ \mathcal{V}_{+} \mathcal{V}_{-} \mathcal{V}_0 
+ \mathcal{V}_{+} \mathcal{V}_0 \mathcal{V}_{-}
+ \mathcal{V}_0 \mathcal{V}_{+} \mathcal{V}_{-} 
\right)
\nonumber \\
& & 
+ \frac{1}{3!} 
\left(
  \mathcal{V}_{-}^3
+ \mathcal{V}_{-}^2 \mathcal{V}_{+} 
+ \mathcal{V}_{-}^2 \mathcal{V}_0 
+ \mathcal{V}_0 \mathcal{V}_{-} \mathcal{V}_{0}
+ \mathcal{V}_{-} \mathcal{V}_0^2
+ \mathcal{V}_0^2 \mathcal{V}_{-}
+ \mathcal{V}_0 \mathcal{V}_{-}^2
+ \mathcal{V}_{-} \mathcal{V}_0 \mathcal{V}_{-}
+ \mathcal{V}_{-} \mathcal{V}_{+} \mathcal{V}_{-}
+ \mathcal{V}_{+} \mathcal{V}_{-}^2
\right).
\nonumber \\
\end{eqnarray}
On the other hand, the right hand side in \eqref{eq:cubic} is 
\begin{eqnarray}
& & e^{\mathcal{V}_{-} + f^{(2)}_{-} + f^{(3)}_{-}}
 e^{\mathcal{V}_0 + f^{(2)}_0 + f^{(3)}_0} e^{\mathcal{V}_{+} +
 f^{(2)}_{+} + f^{(3)}_{+}}|_{\mathcal{O} (\mathcal{V}^3)} 
\nonumber \\
& & 
= 
\left(
\frac{1}{3!} 
\mathcal{V}_{+}^3 + f^{(3)}_{+} 
+ \frac{1}{2} \mathcal{V}_{+} f^{(2)}_{+} 
+ \frac{1}{2} f^{(2)}_{+} \mathcal{V}_{+}
+ \mathcal{V}_0 f^{(2)}_{+} + f^{(2)}_0 \mathcal{V}_{+}
+ \frac{1}{2} \mathcal{V}_0 \mathcal{V}_{+} 
+ \frac{1}{2} \mathcal{V}_{-} \mathcal{V}_{+}^2
+ \frac{1}{2} \mathcal{V}_0 \mathcal{V}_{+}^2
\right)
\nonumber \\
& & 
+ 
\left(
\frac{1}{3!} 
\mathcal{V}_{0}^3 + f^{(3)}_0 + \frac{1}{2} \mathcal{V}_0 f^{(2)}_0 
+ \frac{1}{2} f^{(2)}_0 \mathcal{V}_0 
+ f^{(2)}_{-} \mathcal{V}_{+} + \mathcal{V}_{-} \mathcal{V}_{0}
\mathcal{V}_{+}
+ \mathcal{V}_{-} f^{(2)}_{+}
\right), \nonumber \\
& & 
+
\left(
\frac{1}{3!} 
\mathcal{V}_{-}^3 
+ f^{(3)}_{-} + \frac{1}{2} \mathcal{V}_{-} f^{(2)}_{-}
+\frac{1}{2} f^{(2)}_{-} \mathcal{V}_{-}
+ f^{(2)}_{-} \mathcal{V}_{0} + \mathcal{V}_{-} f^{(2)}_{0}
+ \frac{1}{2} \mathcal{V}_{-} \mathcal{V}_{0}^2
+ \frac{1}{2} \mathcal{V}_{-}^2 \mathcal{V}_{+}
+ \frac{1}{2} \mathcal{V}_{-}^2 \mathcal{V}_{0}
\right).
\nonumber \\
\end{eqnarray}
From these results, the functions $f^{(3)}_{0}, f^{(3)}_{\pm}$ are
determined to be 
\begin{eqnarray}
f^{(3)}_{+} &=& 
\frac{1}{12} \zeta^2 
\left[
\frac{}{}
V_1 [V_0, V_1] + [V_1,V_0] V_1
\right]
\nonumber \\
& & 
+
\frac{1}{6} \zeta
\left[
\frac{}{}
V_1 [V_1,V_{-1}] + [V_{-1}, V_1] V_1 + V_0 [V_0, V_1] + [V_1, V_0] V_1
\right], \\
f^{(3)}_0 &=& - \frac{1}{12}
\left[
\frac{}{}
[V_0,V_{-1}] V_1 + V_{-1} [V_1,V_0] + [V_0,V_1] V_{-1} + V_1 [V_{-1}, V_0]
\right], \\
f^{(3)}_{-} &=& 
\frac{1}{12} \zeta^{-2} 
\left[
\frac{}{}
V_{-1} [V_0, V_{-1}] + [V_{-1}, V_0] V_{-1}
\right]
\nonumber \\
& & 
+ \frac{1}{6} \zeta^{-1}
\left[
\frac{}{}
V_{-1} [V_{-1}, V_1] + [V_1, V_{-1}] V_{-1} + V_0 [V_0, V_{-1}] +
[V_{-1}, V_0] V_0
\right].
\end{eqnarray}
Generalizations to higher orders $\mathcal{O} (V^n) \ (n \ge 4)$ are straightforward. 
By using the functions $f^{(n-1)}_{\pm}, f^{(n-1)}_0$ obtained at order $\mathcal{O}
(V^{n-1})$, the equations that determine $f^{(n)}_{\pm}, f^{(n)}_0$
become linear. We can easily solve the equations and find the functions
at order $\mathcal{O} (V^n)$ without any ambiguities.

\section{Anti-commutation relations of gauge covariant derivatives}
The $\mathcal{N} = 3$ gauge covariantized supercovariant derivatives satisfy the following
anti-commutation relations,
\begin{eqnarray}
\begin{aligned}
 & \{\mathcal{D}_{\alpha}, \mathcal{D}_{\beta} \} 
= \mathbb{D}_{(\alpha} \Gamma^{(-)}_{-2 \beta )} + 
\{
\Gamma^{(-)}_{-2 \alpha}, \Gamma^{(-)}_{-2 \beta}
\}, \\
 & \{\mathcal{D}_{\alpha}, \mathcal{D}^{12}_{\beta}\} 
= - \frac{1}{2} \mathbb{D}_{\alpha} \Gamma^{(-)}_{-1 \beta} +
D^{12}_{\beta} \Gamma^{(-)}_{-2 \alpha} - \frac{1}{2} 
\{
\Gamma^{(-)}_{-2 \alpha}, \Gamma^{(-)}_{-1 \beta}
\}, \\
 & \{\mathcal{D}_{\alpha}, \bar{\mathcal{D}}_{\beta}\}
= \{ \mathbb{D}_{\alpha}, \bar{\mathbb{D}}_{\beta}\} 
- \mathbb{D}_{\alpha} \Gamma^{(-)}_{0 \beta} + \bar{\mathbb{D}}_{\beta}
\Gamma^{(-)}_{-2 \alpha} - \{ \Gamma^{(-)}_{-2 \alpha}, \Gamma^{(-)}_{0
\beta} \}, \\
 & \{\mathcal{D}^{12}_{\alpha}, \mathcal{D}^{12}_{\beta}\} 
= \{D^{12}_{\alpha},D^{12}_{\beta}\}
- \frac{1}{2} D^{12}_{(\alpha} \Gamma^{(-)}_{-1 \beta)} + \frac{1}{4}
\{\Gamma^{(-)}_{-1 \alpha}, \Gamma^{(-)}_{-1 \beta}\}, \\
 & \{\mathcal{D}^{12}_{\alpha}, \bar{\mathcal{D}}_{\beta} \} 
= - D^{12}_{\alpha} \Gamma^{(-)}_{0 \beta} 
- \frac{1}{2}
\bar{\mathbb{D}}_{\beta} \Gamma^{(-)}_{-1 \alpha} + \frac{1}{2} 
\{
\Gamma^{(-)}_{-1 \alpha}, \Gamma^{(-)}_{0 \beta}
\}, \\
 & \{\bar{\mathcal{D}}_{\alpha}, \bar{\mathcal{D}}_{\beta}\}
= - \bar{\mathbb{D}}_{(\alpha} \Gamma^{(-)}_{0 \beta)} + 
\{
\Gamma^{(-)}_{0 \alpha}, \Gamma^{(-)}_{0 \beta}
\}.
\end{aligned}
\end{eqnarray}
The anti-commutation relations of the $\mathcal{N} = 4$ 
gauge covariantized supercovariant derivatives in the left part are 
\begin{eqnarray}
\begin{aligned}
 & \{\mathcal{D}^{1 \bar{1}}_{\alpha}, \mathcal{D}^{1 \bar{1}}_{\beta}
 \} =  0, \\
 & \{\mathcal{D}^{1 \bar{1}}_{\alpha}, \mathcal{D}^{1 \bar{2}}_{\beta}
 \} =  0, \\
 & \{\mathcal{D}^{1 \bar{1}}_{\alpha}, \mathcal{D}^{2 \bar{1}}_{\beta}
 \} =  \mathbb{D}_{\alpha} \Gamma^{(-) \bar{1}}_{0 \beta}, \\
 & \{\mathcal{D}^{1 \bar{1}}_{\alpha}, \mathcal{D}^{2 \bar{2}}_{\beta}
 \} = \{D^{1 \bar{1}}_{\alpha}, D^{2 \bar{2}}_{\beta} \} +
 \mathbb{D}_{\alpha} \Gamma^{(-) \bar{2}}_{0 \beta}, \\
 & \{\mathcal{D}^{1 \bar{2}}_{\alpha}, \mathcal{D}^{1 \bar{2}}_{\beta}
 \} = 0, \\
 & \{\mathcal{D}^{1 \bar{2}}_{\alpha}, \mathcal{D}^{2 \bar{1}}_{\beta}
 \} = \{D^{1 \bar{2}}_{\alpha}, D^{2 \bar{1}}_{\beta} \}
+ D^{1\bar{2}}_{\alpha} \Gamma^{(-) \bar{1}}_{0 \beta}, \\
 & \{\mathcal{D}^{1 \bar{2}}_{\alpha}, \mathcal{D}^{2 \bar{2}}_{\beta}
 \} = D^{1 \bar{2}}_{\alpha} \Gamma^{(-) \bar{2}}_{0 \beta} , \\
 & \{\mathcal{D}^{2 \bar{1}}_{\alpha}, \mathcal{D}^{2 \bar{1}}_{\beta}
 \} = D^{2 \bar{1}}_{\alpha} \Gamma^{(-) \bar{1}}_{0 \beta} 
+ D^{2 \bar{1}}_{\beta} \Gamma^{(-) \bar{1}}_{0 \alpha} 
+ \{\Gamma^{(-) \bar{1}}_{0 \alpha}, \Gamma^{(-) \bar{1}}_{0 \beta} \}, \\
 & \{\mathcal{D}^{2 \bar{1}}_{\alpha}, \mathcal{D}^{2 \bar{2}}_{\beta}
 \} = D^{2 \bar{1}}_{\alpha} \Gamma^{(-) \bar{2}}_{0 \beta} 
- \bar{\mathbb{D}}_{\beta} \Gamma^{(-) \bar{1}}_{0 \alpha} + 
\{\Gamma^{(-) \bar{1}}_{0 \alpha}, \Gamma^{(-) \bar{2}}_{0 \beta} \}, \\
 & \{\mathcal{D}^{2 \bar{2}}_{\alpha}, \mathcal{D}^{2 \bar{2}}_{\beta}
 \} = - \bar{\mathbb{D}}_{\alpha} \Gamma^{(-) \bar{2}}_{0 \beta} -
 \bar{\mathbb{D}}_{\beta} \Gamma^{(-)\bar{2}}_{0 \alpha} + \{\Gamma^{(-)
 \bar{2}}_{0 \alpha}, \Gamma^{(-) \bar{2}}_{0 \beta} \} .
\end{aligned}
\end{eqnarray}
Similar relations hold in the right sector.

In four dimensions, the $\mathcal{N} = 2$ gauge covariantized
supercovariant derivatives satisfy the following algebras,
\begin{eqnarray}
\begin{aligned}
 & \{\mathcal{D}_{1 \alpha}, \mathcal{D}_{1 \beta} \} = 
D_{1 \alpha} \Gamma^{(-)}_{0 \beta} + D_{1 \beta} \Gamma^{(-)}_{0
\alpha} + \{ \Gamma^{(-)}_{0 \alpha}, \Gamma^{(-)}_{0 \beta} \}, \\
 & \{\mathcal{D}_{1 \alpha}, \mathcal{D}_{2 \beta} \} = 
D_{2 \beta} \Gamma^{(-)}_{0 \alpha}, \\
 & \{\mathcal{D}_{1 \alpha}, \bar{\mathcal{D}}^1_{\dot{\alpha}} \} = 
\{D_{1 \alpha}, \bar{D}^1_{\dot{\alpha}}\} + \bar{D}^1_{\dot{\beta}}
\Gamma^{(-)}_{0 \alpha}, \\
 & \{\mathcal{D}_{1 \alpha}, \bar{\mathcal{D}}^2_{\dot{\alpha}} \} = 
D_{1 \alpha} \bar{\Gamma}^{(-)}_{0 \dot{\alpha}} +
\bar{D}^2_{\dot{\beta}} \Gamma^{(-)}_{0 \alpha} + 
\{\Gamma^{(-)}_{0 \alpha}, \Gamma^{(-)}_{0 \dot{\alpha}}\}, \\
 & \{\mathcal{D}_{1 \alpha}, \mathcal{D}_{2 \beta} \} = 0, \\
 & \{\mathcal{D}_{1 \alpha}, \bar{\mathcal{D}}^1_{\dot{\alpha}} \} =
 0, \\
 & \{\mathcal{D}_{1 \alpha}, \bar{\mathcal{D}}^2_{\dot{\alpha}} \} = 
\{D_{2 \alpha}, \bar{D}^2_{\dot{\alpha}} \} + D_{2 \alpha}
\bar{\Gamma}^{(-)}_{0 \dot{\alpha}}, \\
 & \{ \bar{\mathcal{D}}^1_{\dot{\alpha}},
 \bar{\mathcal{D}}^1_{\dot{\beta}} \} = 0, \\
 & \{ \bar{\mathcal{D}}^1_{\dot{\alpha}},
 \bar{\mathcal{D}}^2_{\dot{\beta}} \} = \bar{D}^1_{\dot{\alpha}}
 \bar{\Gamma}^{(-)}_{0 \dot{\beta}}, \\
 & \{ \bar{\mathcal{D}}^2_{\dot{\alpha}},
 \bar{\mathcal{D}}^2_{\dot{\beta}} \} = 
\bar{D}^2_{\dot{\alpha}} \bar{\Gamma}^{(-)}_{0 \dot{\beta}} +
\bar{D}^2_{\dot{\beta}} \Gamma^{(-)}_{0 \dot{\alpha}} + 
\{\bar{\Gamma}^{(-)}_{0 \dot{\alpha}},
 \bar{\Gamma}^{(-)}_{0\dot{\beta}}\}.
\end{aligned}
\end{eqnarray}

\end{document}